\documentclass[twocolumn,prb,superscriptaddress,floatfix,showpacs]{revtex4}

\usepackage{graphicx,latexsym,dcolumn}

\newcommand{\ctd}{\ensuremath{C_{\text{2D}}}}
\newcommand{\ceff}{\ensuremath{C_{\text{eff}}}}
\newcommand{\co}{\ensuremath{C_{1}}}
\newcommand{\ct}{\ensuremath{C_{2}}}
\newcommand{\csig}{\ensuremath{C_{\Sigma}}}
\newcommand{\cg}{\ensuremath{C_{g}}}

\newcommand{\ec}{\ensuremath{E_{c}}}

\newcommand{\rl}{\ensuremath{r_{\ell}}}
\newcommand{\rli}{\ensuremath{r_{\ell_{i}}}}
\newcommand{\cl}{\ensuremath{c_{\ell}}}
\newcommand{\cli}{\ensuremath{c_{\ell_{i}}}}
\newcommand{\iv}{$I$-$V$}
\newcommand{\rqu}{\ensuremath{R_{Q}}}
\newcommand{\rk}{\ensuremath{R_{K}}}
\newcommand{\kb}{\ensuremath{k_{B}}}
\newcommand{\rsq}{\ensuremath{R_{sq}}}
\newcommand{\ztd}{\ensuremath{Z_{\text{2D}}}}
\newcommand{\zl}{\ensuremath{Z_{\ell}}}
\newcommand{\zrc}{\ensuremath{Z_{RC}}}
\newcommand{\zt}{\ensuremath{Z_{t}}}
\newcommand{\rezt}{\ensuremath{\text{Re}[\zt(\omega)]}}

\newcommand{\rqpc}{\ensuremath{R_{\text{QPC}}}}
\newcommand{\rstr}{\ensuremath{R_{\text{str}}}}
\newcommand{\cstr}{\ensuremath{C_{\text{str}}}}
\newcommand{\zstr}{\ensuremath{Z_{\text{str}}}}
\newcommand{\tstr}{\ensuremath{\tau_{\text{str}}}}
\newcommand{\gset}{\ensuremath{G_{\text{SET}}}}
\newcommand{\gsc}{\ensuremath{\gset^{c}}}

\newcommand{\gtd}{\ensuremath{G_{\text{2D}}}}
\newcommand{\sch}{\ensuremath{\sigma_{ch}}}
\newcommand{\vg}{\ensuremath{V_{g}}}

\newcommand{\fc}{\ensuremath{{\cal F}_{c}}}
\newcommand{\fs}{\ensuremath{{\cal F}_{s}}}
\newcommand{\units}[1]{\ensuremath{\mathrm{#1}}}
\newcommand{\amount}[2]{\ensuremath{#1~\units{#2}}}
\newcommand{\e}[1]{\ensuremath{\times 10^{#1}}}
\newcommand{\etal}{\textit{et al.}}
\newcommand{\ie}{\textit{i.~e.}}
\newcommand{\alxgas}{GaAs/Al$_{x}$Ga$_{1-x}$As}
\newcommand{\alx}{\ensuremath{{\rm Al/AlO}_{x}}}
\newcommand{\tfrac}[2]{\ensuremath{{\textstyle\frac{#1}{#2}}}}
\newcommand{\half}{\tfrac{1}{2}}
\newcommand{\ehz}{\ensuremath{e/\sqrt{\mathrm{Hz}}}}

\begin{document}

\title{Superconducting single-electron transistor coupled to a two dimensional electron gas: Transmission lines, dissipation, and charge averaging}
\author{A. J. Rimberg}
\affiliation{Department of Physics and Astronomy, Rice University,
Houston, Texas 77005}
\affiliation{Department of Electrical and Computer Engineering, Rice University, 
Houston, Texas 77005}
\author{W. Lu}
\affiliation{Department of Physics and Astronomy, Rice University,
Houston, Texas 77005}

\begin{abstract}
We have developed a novel system consisting of a superconducting
single-electron transistor (S-SET) coupled to a two-dimensional electron
gas (2DEG), for which the dissipation can be tuned in the immediate
vicinity of the S-SET\@. To analyze our results, we have developed a
model of the environment for S-SET/2DEG systems that includes
electromagnetic fluctuations coupled both through the S-SET leads and
capacitively to the S-SET central island.  We analyze this model,
treating the leads as finite transmission lines, to find the probability
function $P(E)$ for exchanging energy $E$ with the environment.   We
also allow for the possibility of low-frequency fluctuations of the S-SET
offset charge.  We compare our calculations with measurements of SET
conductance versus 2DEG conductance and find good agreement for
temperatures $>\amount{100}{mK}$, while unexplained discrepancies emerge
for lower temperatures.  By including the effects of charge averaging we
are also able to predict the shape and evolution of \iv\ curves as the
2DEG in the vicinity of the S-SET is changed.
\end{abstract}


\maketitle

\section{Introduction}

The effects of the electromagnetic environment on electric transport has
been a subject of extensive theoretical and experimental interest in
recent years.  The reasons for interest are varied, as are the systems
for which studies of the effects of the environment have been performed.
Recent interest in quantum
computation\cite{Makhlin:1999,Nakamura:1999,Mooij:1999,Vion:2002} has
prompted interest in the effects of dissipation on decoherence rates in
superconducting qubits.\cite{Shnirman:1997,Makhlin:2001a} Double quantum
dots have been used to study the effects of the environment on inelastic
tunneling rates,\cite{Fujisawa:1998} and have been proposed as detectors
of high-frequency noise produced by mesoscopic
devices.\cite{Aguado:2000} Finally, interest in quantum phase
transitions\cite{Sondhi:1997} has prompted study of the effects of
dissipation on superconducting systems such as thin
films\cite{Mason:1999,Kapitulnik:2001} and Josephson junction
arrays.\cite{Rimberg:1997}

Given the possibility of using a single Cooper pair
box\cite{Bouchiat:1998,Nakamura:1999} as a qubit and its similarity to
the superconducting single-electron transistor
(S-SET),\cite{Grabert:1992} it seems logical to use the S-SET as a model
system for studying the effects of environmental dissipation on
coherence and transport in small tunnel junction systems.  Recently,
there have been several such attempts, motivated by experiments in which
a Josephson junction array was fabricated in close proximity to a
two-dimensional electron gas (2DEG) in an \alxgas\
heterostructure,\cite{Rimberg:1997} which can be used as a tunable
source of dissipation.  (While the effects of a mechanically tunable
environment were earlier studied in the macroscopic quantum tunneling
regime,\cite{Turlot:1989} use of a 2DEG allows more flexible tuning of
the environment over a larger impedance range.) In place of an array, we
and the Berkeley group have instead used similar fabrication techniques
to couple an S-SET to a 2DEG\@.\cite{Kycia:2001,Lu:2002,Lu:2002a} In one
instance, the focus was on transport at higher biases, in the regime of
the Josephson-quasiparticle cycle.\cite{Lu:2002}  In the others, the
focus instead was on the low bias regime and the tunneling rates of
Cooper pairs.\cite{Kycia:2001,Lu:2002a}  Theoretical work aimed at an
explanation of the results of the Berkeley group was undertaken by
Wilhelm, \etal\cite{Wilhelm:2001}

The primary experimental difference between our own work and that of the
Berkeley group lies in the way in which the environment is varied.  The
Berkeley group followed an approach developed earlier for the study of
junction arrays,\cite{Rimberg:1997} using a gate on the back side of
the substrate to vary the sheet density $n_{s}$ of the 2DEG, and
therefore its resistance per square \rsq.  Such a change is global, and
affects the 2DEG not only immediately beneath the S-SET, but
beneath the macroscopic leads used to measure it as well.  In our own
work, by placing Au gates on the surface of the sample near the S-SET
itself, we were able to vary the dissipation in the 2DEG locally, while
leaving 2DEG beneath the leads virtually unchanged.\cite{Lu:2000,Lu:2002a}

The Berkeley group compared their results to the theory of Wilhelm,
\etal, which predicted that within linear response the conductance
\gset\ of the S-SET would scale with the ground plane conductance
$\gtd=1/\rsq$ and temperature $T$ as $\gtd^{\beta}/T^{\alpha}$. 
While the Berkeley group did observe power law behavior, their
measured exponents were not in quantitative agreement with theory. 
Furthermore, the measured $\beta$ depended on $T$ and $\alpha$ on \gtd,
calling the scaling form into question.  

In our own work,\cite{Lu:2002a} we examined a somewhat more complex model for the
environment than that considered by Wilhelm, \etal\  Specifically, we
allowed for coupling of electromagnetic fluctuations to tunneling Cooper
pairs due simultaneously both to the S-SET leads and to the 2DEG in
the immediate vicinity of the S-SET, which is coupled to the S-SET central
island via a capacitance \ctd.  By also allowing for averaging of the
S-SET offset charge, we were able to obtain good agreement between our
measurements and calculations.   Here we examine our model of the
environment in somewhat more detail, provide additional experimental
data which supports our earlier analysis, and also give some additional
details of the calculation.

\section{Sample and Environmental Calculations}  

\subsection{Sample Design and the Environmental Model}

\subsubsection{Sample Design}

Our samples consist of an \alx-based S-SET fabricated in close proximity
to a 2DEG formed in a \alxgas\ heterostructure\cite{Lu:2000} as shown in
Fig.~\ref{sample} below.  We begin by fabricating six Au gates which can
be used to deplete the electrons beneath them by application of a
negative gate voltage \vg.  At the center of the Au gates we then
fabricate our S-SET, as can be seen in the electron micrograph (expanded view
in Fig.~\ref{sample}).  Note that as shown in the larger diagram the
S-SET leads extend over the 2DEG to macroscopic contact pads. For the
vast majority of their length they are well away from the Au gates and
the 2DEG beneath them is independent of \vg.  When we apply a gate
voltage \vg\ to all six Au gates, the electrons immediately beneath them
are depleted, leaving a small pool of electrons beneath the S-SET\@. 
This pool is connected to the rest of the 2DEG (held at ground) only by
two quantum point contacts (QPCs) with conductances $1/\rqpc$ (assumed
equal) as shown in the micrograph.  It is also capacitively coupled to
the S-SET island through a capacitance \ctd\ as shown in the lower right
inset to Fig.~\ref{sample}.  When all six Au gates are energized as
described above, we say that the electrons are confined in the ``pool''
geometry.  We do not refer to the pool as a quantum dot since for these
experiments the QPCs are sufficiently open that no Coulomb oscillations
are detected in the pool and discrete energy levels have not formed.

Because the Au gates can be biased independently, we can also apply \vg\
to only the four outermost gates which form the QPCs.  As before, the
electrons beneath the S-SET are coupled to ground through the QPCs. In
addition, however, they are now coupled through a resistance \rstr\ to
two large reservoirs of electrons located between the four outermost
gates, as can be seen in Fig.~\ref{sample}.  The reservoirs are in turn
coupled to ground only through a capacitance \cstr.   When only the four
outer gates are energized, we say that the electrons in the 2DEG are
confined in the ``stripe'' geometry.  We observe significant differences
between the measured S-SET conductance \gset\ versus applied gate
voltage \vg\ for the two different geometries, as will be discussed below.
\begin{figure}[b!]
\includegraphics[width=2.9in]{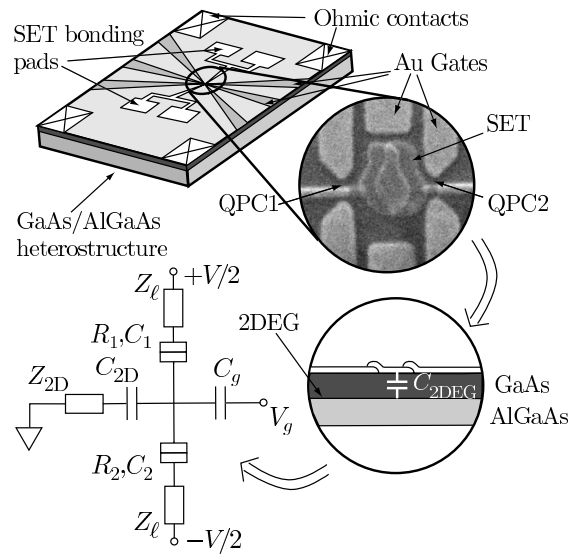}
\caption{\label{sample} Schematic diagram of a typical sample, showing
the Au gates, Ohmic 2DEG contacts, and SET leads and bonding pads.  The
upper right inset is an electron micrograph showing the S-SET island
surrounded by the six Au gates.  Application of a gate voltage \vg\
causes QPCs to form at the locations shown.  The S-SET is coupled to the
2DEG beneath it by a capacitance \ctd\ as illustrated in the lower right
inset.  In the lower left we show a circuit diagram of our model of the
S-SET environment, including the lead impedances \zl\ and the impedance
\ztd\ associated with the electrons immediately beneath the S-SET,
coupled to it through the capacitance \ctd.  We also show the combined
capacitance \cg\ to the six Au gates.}
\end{figure}

Regardless of the gate configuration used, we can apply a single model
of the environment to our results, as shown in the lower left in
Fig.~\ref{sample}.  The SET island is connected to its leads through
junctions with resistance $R_{1(2)}$ and capacitance $C_{1(2)}$.   We
assume that the S-SET leads present an impedance \zl\ to the SET while
the 2DEG electrons have a total impedance \ztd\ to ground which is
coupled to the SET through the capacitance \ctd.  Nearly the entire
length of the SET leads is far from the Au gates, so that \zl\ is almost
completely unaffected by the gate voltage \vg.  The electrons
immediately beneath the SET are strongly affected by \vg\, so that \ztd\
will in general be a function of \vg, and may also depend on the
configuration of gates used ({\itshape i.\ e.}, on the pool versus
stripe geometry). Finally, the SET is also coupled to the Au gates by a
capacitance \cg.  We neglect the possibility of a substantial impedance
on the gate lines largely because $\cg\approx\amount{20}{aF}$ is by far
the smallest capacitance in the problem.  Furthermore, any gate
impedance would be substantially reduced since there are six gates whose
impedances would combine in parallel. This general model (excluding the
small gate capacitance \cg) has been investigated
previously,\cite{Ingold:1991,Odintsov:1991} but without considering any
particular form for the impedances \zl\ and \ztd.

In Table~\ref{params} below we give the relevant sample parameters for
the two samples S1 and S2 considered here.  The parameters were
determined from electrical measurement and simulations as discussed
elsewhere.\cite{Lu:2002}

\begin{table}
\caption{\label{params}Parameters for samples S1 and S2.  Capacitances are
in aF and energies in $\mu$eV.}
\begin{ruledtabular}
\begin{tabular}{cccccrdd}
sample & \co & \ct & \cg & \ctd & \multicolumn{1}{c}{\ec} &%
 \multicolumn{1}{c}{$E_{J_{1}}$} & \multicolumn{1}{c}{$E_{J_{2}}$}\\ \hline
S1 & 181 & 120 & 20 & 356 & 118 & 5.9 & 3.5 \\
S2 & 375 & 260 & 20 & 382 & 77 & 27. & 16. \\
\end{tabular}
\end{ruledtabular}
\end{table}

\subsubsection{Tunneling Rates}
In general for our samples the charging energy $\ec=e^2/2\csig$ where
$\csig=\co+\ct+\ctd+\cg$ satisfies $E_{J_{j}}<\ec\ll\kb T$, where
$E_{J_{j}}=\frac{\rqu}{2R_{j}}\Delta$ is the Josephson energy of
junction $j$ given by the Ambegaokar-Baratoff
relation\cite{Ambegaokar:1963} and $\Delta$ is the superconducting gap. 
Under these circumstances the S-SET island charge is well defined, so
that charge states can be used as the basis for calculating the
tunneling rates.\cite{Ingold:1992,Schon:1998a}  We will also be
concerned with transport at sufficiently low bias voltages $V$ and
temperatures $T$, that we need only consider the tunneling of Cooper
pairs, for which the sequential tunneling rate through junction $j$ is
given by\cite{Averin:1990}
\begin{equation}\label{gameq}
\Gamma(\delta f^{(j)}) =
(\pi/2\hbar) E_{J_{j}}^{2}P(-\delta f^{(j)}) 
\end{equation}
which is valid for $E_{J_{j}}\ll\ec$.  Here $\delta f^{(j)} = f_{f} -
f_{i}$ is the change in free energy associated with the tunneling event,
and the function $P(E)$ describes the probability of the Cooper pair
exchanging an energy $E$ with the electromagnetic environment during the
tunneling process.  Following the usual environmental
theory,\cite{Ingold:1992} $P(E)$ can be expressed in terms of the real
part of the total impedance seen by the tunneling electron
\rezt\ first through a kernel $K(t)$
\begin{eqnarray}\label{koteq}
\lefteqn{K(t) = \rqu^{-1}\int_{-\infty}^{\infty}\frac{d\omega}{\omega}
\rezt}\qquad\nonumber \\ 
 & & \times\{\coth(\frac{\hbar \omega}{2\kb T})[\cos(\omega
t)-1] - i \sin(\omega t)\}
\end{eqnarray}
and then through the Fourier transform 
\begin{equation}\label{peeq}
P(E)=\frac{1}{2\pi\hbar}\int_{-\infty}^{\infty}dt\;\exp[K(t)+iEt/\hbar].
\end{equation}
A calculation of the tunneling rate $\Gamma$ must therefore begin with a
clear understanding of the impedance $\zt(\omega)$ presented to the
tunneling electrons by the environment.
 
\subsubsection{Model of the Environment}
 
Given the circuit model shown in Fig.~\ref{sample}, one can use standard
network analysis\cite{Grabert:1991} to calculate the impedance
$\zt(\omega)$ seen by an electron tunneling through junction $j$ in
terms of the impedances and capacitances shown in Fig.~\ref{sample}. The
result is given by\cite{Ingold:1991}
\begin{equation}\label{zteq}
\zt(\omega) = \frac{1}{i\omega \widetilde{C} + \widetilde{Y}}
\end{equation}
where
\begin{equation}\label{cteq}
\widetilde{C} = \frac{\csig C_{j}}{C_{j'}+\ctd}
\end{equation}
where $j'= 2(1)$ for $j=1(2)$ and 
\begin{widetext}
\begin{equation}\label{yteq}
\widetilde{Y} = \frac{\csig^2}{(C_{j'}+\ctd)C_{j'}\ctd}\;%
\frac{(C_{j'}+\ctd)C_{j'}\ctd +i\omega(C_{j'}\ctd)^2(\zl+\ztd)}{[(C_{j'}%
+\ctd)^2+C_{j'}^2]\zl + \ctd^2\ztd + 
i\omega(C_{j'}+\ctd)C_{j'}\ctd(\zl+2\ztd)\zl}
\end{equation}
\end{widetext}
ignoring terms of order $\cg/\ctd$.  To proceed, we need accurate models
of \zl\ and \ztd; we begin by considering \zl.

\subsubsection{Model of \zl}

Since our leads are fabricated above the 2DEG, which acts as a ground
plane, it is appropriate to model them as transmission
lines.\cite{Pozar:1998} The most general form for the impedance
$Z_{\text{tr}}$ of a lossy transmission line terminated in a load $Z_L$
is given by
\begin{equation}\label{lleq}
Z_{\text{tr}}=Z_{0}\frac{Z_{L}+Z_{0}\tanh \gamma \ell}{Z_{0}+Z_{L}\tanh \gamma \ell}
\end{equation}
where $Z_0$ is the characteristic impedance of the line, $\gamma$ its
complex propagation constant and $\ell$ its length.  At the relatively
low frequencies ($\alt \amount{10^{11}}{Hz}$) considered here, it is
reasonable to ignore the inductive reactance of the line and treat it as
a simple $RC$ line with a resistance and capacitance per unit length \rl\
and \cl.   Doing so, we have that $Z_{0}=\sqrt{\rl/i\omega\cl}$ and
$\gamma=\sqrt{i\omega\cl\rl}$.  Looking out at the line from the sample,
the line termination $Z_L$ is provided by the bias circuitry, which
typically presents a low impedance $\alt\amount{50}{\Omega}$.  For
simplicity we therefore take $Z_L=0$ in Eq.~(\ref{lleq}), and obtain the
resulting approximation
\begin{equation}\label{zrceq}
\zrc(\omega)=\sqrt{\frac{\rl}{i\omega\cl}}\tanh\sqrt{i\omega\rl\cl\ell^2} 
\end{equation}
which we take as the basic form for the impedance of a finite $RC$ line.
 This form has been considered previously in the context of incoherent
tunneling of Cooper pairs in individual Josephson
junctions.\cite{Kuzmin:1991}

While this is likely a fairly accurate description of the impedance of a
section of our leads, when used in evaluating the kernel $K(t)$ in
(\ref{koteq}) it leads to integrals which are analytically intractable. 
Fortunately a further simplification is possible.  We are interested in
the low energy part of $P(E)$, which we expect from (\ref{peeq}) to be
dominated by the long time behavior of $K(t)$, which is in turn
dominated by the low frequency part of the impedance \zrc.  We therefore
expand Eq.~(\ref{zrceq}) around $\omega=0$ to obtain
\begin{equation}\label{zrclfeq}
\zrc(\omega)\approx\frac{\rl\ell}{1+(\omega\rl\cl\ell^2/\sqrt{6})^2}
\end{equation}
as a reasonable approximation to \zrc\ in the interesting limit.  

Another common treatment\cite{Schon:1998a} of the $RC$ transmission line
problem is to consider an infinite $RC$ line, whose impedance is given
by $Z_{0}=\sqrt{\rl/i\omega\cl}$. Unlike the finite $RC$ line, for which
the impedance \zrc\ approaches a constant $\rl\ell$ at $\omega=0$, the
infinite $RC$ line has a $1/\sqrt{\omega}$ singularity at $\omega=0$
which dominates the long-time limit of $K(t)$ and therefore $P(E)$. The
kernel $K(t)$ for the infinite line, as well as $P(E)$, can be
calculated exactly in the $T=0$ limit.   At non-zero temperatures, a
high-temperature expansion must be performed instead.\cite{Wilhelm:2001}

\subsubsection{Model of \ztd}
Having developed a model for \zl, we now consider a model for \ztd.  The
particular model will depend on the geometry we choose.  For an
unconfined 2DEG, the simplest choice is that \ztd\ is ohmic with an
impedance related to \rsq\ of the 2DEG: $\ztd\approx\rsq/3$.  When the
electrons are confined in the pool geometry, they are coupled to the
remaining 2DEG by two QPCs with conductance $1/\rqpc$ (assumed equal),
which appear in parallel from the vantage point of the SET\@.  There is
likely some shunt capacitance $C_{\text{QPC}}$ as well, but the
associated roll-off frequency $1/\rqpc C_{\text{QPC}}$ is typically large
($\sim\amount{10^{11}}{s^{-1}}$) and we therefore neglect it.  So for
the pool geometry, we take
\begin{equation}\label{reztdpeq}
\ztd = \rqpc/2.
\end{equation}

The stripe geometry is more complex.  Here, in addition to the QPC
conductances, the electrons beneath the SET are coupled to two large
electron reservoirs with resistance \rstr\ located between the outermost
Au gates.  At their narrowest, the reservoirs are \amount{0.6}{\mu m}
wide, but broaden in five sections to a width of \amount{500}{\mu m}. 
Each section contributes roughly $2\rsq$, so that $\rstr \approx 10\rsq
=\amount{200}{\Omega}$.  These reservoirs are in turn coupled to ground
capacitively through a capacitance \cstr, which we estimate from the
size of the reservoirs to be on the order of \amount{0.3}{pF}.  Using
$\ztd^{-1}=2(\rqpc^{-1}+\zstr^{-1})^{-1}$ where $\zstr=\rstr+1/i\omega\cstr$
we find
\begin{equation}\label{reztdseq}
\text{Re}[\ztd]=\frac{\rqpc}{2}\left[\frac{1+\omega^2\tstr^2 \rstr/(\rqpc+\rstr)}{1%
+\omega^2\tstr^2}\right]
\end{equation}  
where $\tstr=\cstr(\rstr+\rqpc)$.  For $\omega^2\tstr^2\ll 1$, then,
$\text{Re}[\ztd]$ approaches $\rqpc/2$, while for $\omega^2\tstr^2\gg
1$,  $\text{Re}[\ztd]$ approaches $\half(\rstr^{-1}+\rqpc^{-1})^{-1}$. 
The imaginary part of \ztd\ in the stripe geometry is nonnegligible only
in the vicinity of $\omega\sim1/\tstr$, so for our purposes we neglect
it. In general then, at low frequencies  $\text{Re}[\ztd]$ is kept
finite by the presence of the QPCs, and at higher frequencies is
dominated by the smaller of $\rqpc/2$ and $\rstr/2$.

\subsubsection{Decomposition of $\zt(\omega)$} 

While the form for $\zt(\omega)$ given by Eqs.~(\ref{zteq})--(\ref{yteq}) is
complete, it is generally too complex to make significant headway in
calculating $K(t)$.  Fortunately, significant simplification is
possible.  For typical values of $\rl\approx
1\e{6}$--\amount{1\e{7}}{\Omega/m} and typical line lengths $\ell\approx
0.5$--\amount{1}{mm}, $\zl\gg\ztd$ for small $\omega$.  In contrast, for
sufficiently large $\omega$, \zrc\ becomes quite small
($\alt\amount{10}{\Omega}$) and the condition $\ztd\gg\zl$ is usually
satisfied.  It then becomes possible to decompose $\widetilde{Y}$ in
Eq.~(\ref{yteq}) into a low $\omega$ part dominated by \zl\ and a high
$\omega$ part dominated by \ztd.

For small $\omega$, as long as $\zl\gg\ztd$, we can safely neglect the
terms in Eq.~(\ref{yteq}) involving \ztd.  Furthermore, for $\omega$ such
that $\omega\ll\min(1/C_{j}\zl,1/\ctd\zl)$ we can ignore  terms
in Eqs.~(\ref{zteq}) and (\ref{yteq}) that depend explicitly on $\omega$. 
Making these simplifications, we have that in the small $\omega$, large
\zl\ limit
\begin{equation}\label{ztlfeq}
\zt(\omega)\approx\widetilde{Y}^{-1}=\frac{(C_{j'} + \ctd)^2 +%
C_{j'}^2}{\csig}\zl\equiv\kappa_{1}\zl.
\end{equation}

For large $\omega$, we drop terms of order $\zl/\ztd$, and find that we
can neglect the explicit frequency dependence in the denominator of
(\ref{yteq}) for $\omega\ll\cl/(C_{j'})^2\rl\sim\amount{1\e{16}}{rad/s}$. 
In contrast, we cannot necessarily neglect the explicit frequency
dependence in the numerator, and find
\begin{equation}
\widetilde{Y}\approx\left(\frac{\csig}{\ctd}\right)^{2}\frac{1}{\ztd}+%
i\omega\frac{\csig^{2}C_{j'}}{(C_{j'}+\ctd)\ctd}.
\end{equation}
We combine this with (\ref{zteq}) to find in this limit
\begin{eqnarray}\label{zthfeq}
\lefteqn{\zt(\omega) = \frac{1}{i\omega(\co+\ct)\csig/\ctd +%
(\csig/\ctd)^{2}\ztd^{-1}}}\qquad\qquad\qquad\qquad\qquad \nonumber \\
& & \equiv\frac{1}{i\omega\ceff + \kappa_{2}^{-1}\ztd^{-1}}.
\end{eqnarray}
Combining this result with (\ref{ztlfeq}), we obtain for the real part of $\zt(\omega)$
\begin{equation}\label{rezteq}
\rezt=%
\kappa_{1}\text{Re}[\zl(\omega)]+\frac{\kappa_{2}\ztd}{1+[\omega\ceff\kappa_{2}\ztd]^2},
\end{equation}
which we take as our basic model for the real part of the impedance seen
by an S-SET fabricated above a 2DEG ground plane.  We believe this model
should be applicable not only to our own system, but to that of the
Berkeley group as well.\cite{Kycia:2001}
\begin{figure}[h!]
\includegraphics[width=2.9in]{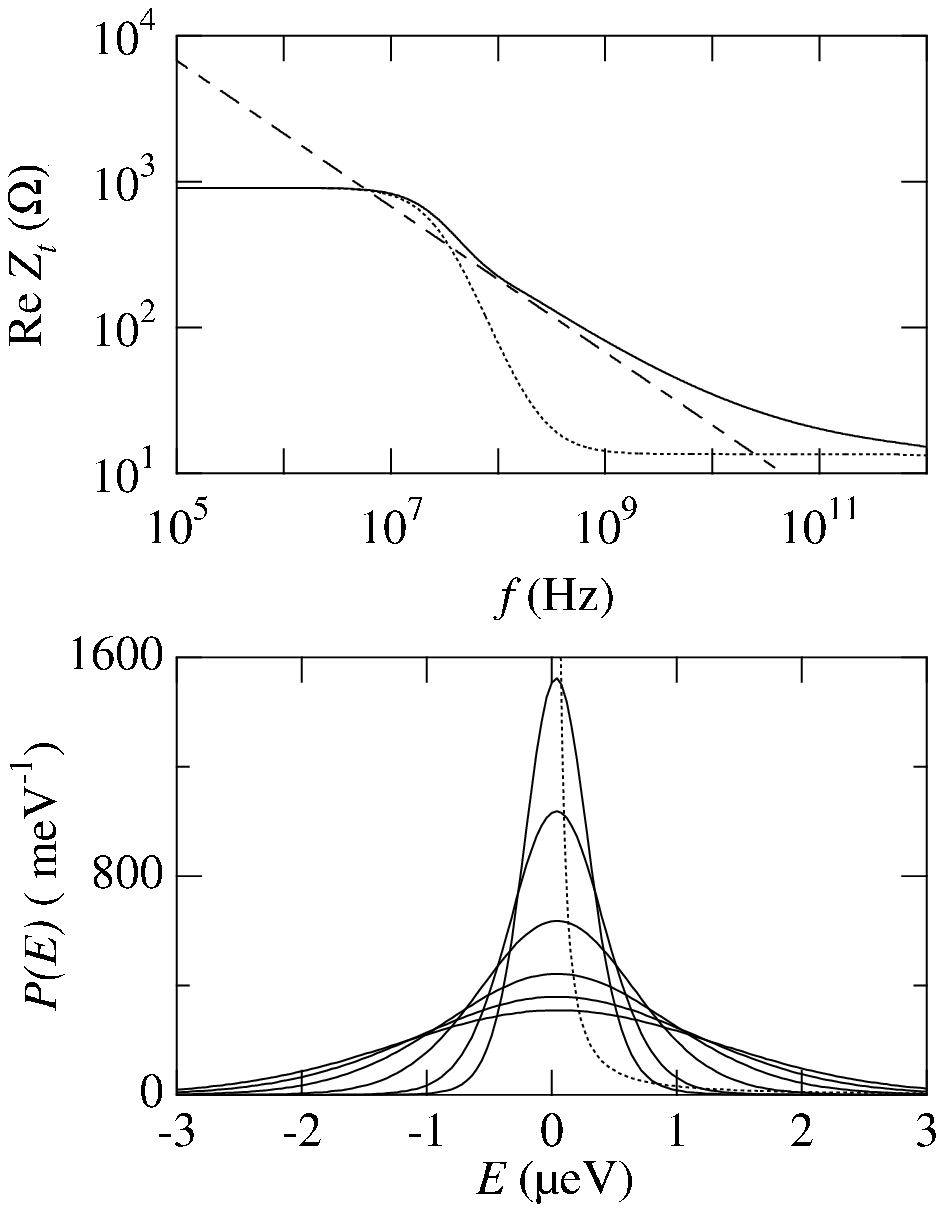}
\caption{\label{zrcfig} (a) \rezt\ for three different environmental
models, all based on a transmission line with
$\rl=\amount{2.9\e{6}}{\omega/m}$ and $\cl=\amount{1.0\e{-8}}{F/m}$ and
$\ell=\amount{6.94\e{-4}}{m}$. For this graph we have used
$\ztd=\amount{100}{\Omega}$. Solid line: exact result based on the full
forms for $\zt(\omega)$ and \zl.  Dotted line: approximate form based on
the low frequency approximation to \zl\ and the decomposition of
\rezt.  Dashed line: infinite line result for the same values
of \rl\ and \cl.  (b) Solid lines: $P_{s}(E)$ for the same transmission
line parameters, for (top to bottom) $T=10$, 20, 50, 100, 150, and
\amount{200}{mK}.  Dashed line: $P(E)$ for an infinite $RC$ line at
$T=0$.}
\end{figure}

To illustrate the degree of approximation associated with
Eqs.~(\ref{zrclfeq}) and (\ref{rezteq}), we show  in
Fig.~\ref{zrcfig}(a) \rezt\ for three separate models of the
environment; all are based on an $RC$ line with
$\rl=\amount{2.9\e{6}}{\omega/m}$ and $\cl=\amount{1\e{-8}}{F/m}$.  The
solid line shows \rezt\ based on Eqs.~(\ref{zteq})--(\ref{yteq}) and
Eq.~(\ref{zrceq}), \ie, on the impedance of a finite $RC$ line coupled
to a ground plane with impedance \ztd, using the full form for \rezt. 
The dotted line is \rezt\ calculated using the low-frequency version of
\zl\ given in (\ref{zrclfeq}) and using the decomposition (\ref{rezteq})
of \rezt, while the dashed line is the impedance of an infinite $RC$
line using the same values of \rl\ and \cl\ (with no ground plane).  We
have not included a curve using the decomposition (\ref{rezteq}) and the
exact form for a finite $RC$ line in (\ref{zrceq}) since it is virtually
indistinguishable from the full \rezt\ shown.

We note that the impedance of the infinite line rises above that of the
more realistic forms for frequencies below a few MHz, so that in general
it gives more weight to the low frequency modes and may be expected to
give a more sharply peaked $P(E)$.  More importantly,  the low-frequency
approximation to \zl, while agreeing quite well below
$\sim\amount{10^7}{Hz}$ with the exact finite line result, significantly
underestimates it for intermediate frequencies below
$\sim\amount{10^{12}}{Hz}$.  The approximation may therefore be of
limited use for larger bias voltages; for low biases, however, it is
likely to be more accurate than a model based on an infinite
transmission line, which overestimates the impedance at low frequencies.
  Finally, at sufficiently high frequencies, the approximate and exact
forms for \rezt\ converge.

\subsection{Calculation of $K(t)$ and $P(E)$}

\subsubsection{Calculation of $K(t)$}

Having produced a tractable form for \rezt, we can now proceed to a
calculation of $K(t)$ and $P(E)$ through Eqs.~(\ref{koteq}) and
(\ref{peeq}).  We begin by noting that when using the low frequency form
for \zl\ in (\ref{zrclfeq}), both parts of \rezt\ have the same form,
namely
\begin{equation}
\rezt=\frac{R}{1 + \omega^2\tau^2}
\end{equation}
for appropriate $R$ and $\tau$. Calculations for $K(t)$ and $P(t)$ for
this form have been given in detail
elsewhere,\cite{Grabert:1998,Grabert:1999} but emphasize a different
range for $R$ and result in different forms for $P(E)$.  Using
\begin{equation}
\frac{1}{\omega}\frac{R}{1 + \omega^2\tau^2}=\frac{R}{\omega}-%
\frac{R\tau^2\omega}{1 +\omega^2\tau^2}
\end{equation}
and $\coth(\hbar \omega/2\kb T)=1 + 2/[\exp(\hbar \omega/\kb T) - 1]$ we find that 
\begin{widetext}
\begin{eqnarray}\label{kotrezteq}
\lefteqn{K(t) = \rqu^{-1}\left\{\fc\left[\frac{R}{\omega}\right]-%
\int\fs\left[\frac{2R}{\exp(\hbar \omega/\kb T)-1}\right]\,dt%
 -i\,\text{sign}(t)\,\fs\left[\frac{1}{\omega}\frac{R}{1 +%
\omega^2\tau^2}\right]\right.}\qquad\qquad\qquad\qquad\qquad\qquad%
\qquad\qquad\nonumber\\
& & \left.\mbox{}-\fc\left[\frac{R\tau^2\omega}{1 +%
\omega^2\tau^2}\coth\left(\frac{\hbar\omega}{2\kb T}\right)\right]\right\}
\end{eqnarray}
\end{widetext}
where $\fc[f(\omega)]=2\int_{0}^{\infty}f(\omega)\cos\omega t\,d\omega$
and $\fs[f(\omega)]=2\int_{0}^{\infty}f(\omega)\sin\omega t\,d\omega$
are the Fourier cosine and sine transforms of $f(\omega)$ respectively,
taken to be functions of $|t|$. We have ignored terms in $K(t)$
independent of $t$; since for $P(E)$ to satisfy the normalization
condition $\int_{-\infty}^{\infty}P(E)\,dE=1$ we must have
$K(0)=0$,\cite{Ingold:1992} we will later ensure normalization by adding
an appropriate constant in any case.

Of the four terms in curly braces in Eq.~(\ref{kotrezteq}) for $K(t)$,
the first three can all be evaluated analytically.\cite{Wilhelm:2001} An
analytic form for the entire kernel has also been
found,\cite{Grabert:1998} and analyzed for the overdamped case such that
$1/\tau$ is large compared to the Josephson frequency
$\omega_{J}=\pi\Delta/\hbar$.  However, the range of $R$ and $\tau$ in
which we are interested was not investigated.  Nevertheless, we have
made some progress in certain limits. We note first that the fourth term
in (\ref{kotrezteq}) depends on the temperature $T$ only through the
dimensionless combination ${\cal T}=\hbar/(2\kb T\tau)$, and write
\begin{equation}
\fc\left[\frac{\tau^2\omega}{1 + %
\omega^2\tau^2}\coth\left(\frac{\hbar\omega}{2\kb T}\right)\right]%
= k_{{\cal T}}(t,{\cal T}) 
\end{equation}
For zero temperature (${\cal T}=\infty$), we have 
\begin{equation}
k_{{\cal T}}(t,\infty)=\sqrt{\pi}G_{1\,3}^{2\,1}\left(\frac{t^2}{4%
\tau^2}\left|\begin{array}{l}0\\0,0,\half\end{array} \right.\right)
\end{equation}
where $G$ is a Meijer G function.  In the long time limit, this result
goes as $-2(\tau/t)^2$.

More generally, for $T\neq 0$, we find that it is important to consider
the relative importance of the terms in Eq.~(\ref{kotrezteq}).  If we
evaluate the integrals which we can treat analytically, we have
\begin{eqnarray}\label{kotaneq}
\lefteqn{K(t)=-\frac{2R}{\rqu}\bigg\{\frac{\pi\kb T |t|}{\hbar}%
+ \ln(1-e^{-2\pi\kb T|t|/\hbar}) + \gamma}\nonumber \\
& &\mbox{}  -\ln 2 + i \frac{\pi}{2}\text{sign}(t)[1 - e^{-|t|/\tau}]%
+\half k_{{\cal T}}(t,{\cal T})\bigg\}
\end{eqnarray}
where $\gamma\approx0.577216$ is Euler's constant.  In order to compare
the relative size of the terms, we evaluate $k_{{\cal T}}$ numerically. 
We find that for ${\cal T}\gg 1$, (either low temperature or small
$\tau$), $k_{{\cal T}}$ decays slowly with time.  For long times then
the term going as $e^{-|t|/\tau}$ is by far the smallest, and can be
neglected. Of the remaining terms, in the long time limit the
logarithmic term dominates over $k_{{\cal T}}$ and we write the kernel
as
\begin{eqnarray}\label{kleq}
\lefteqn{K_{l}(t)=-\frac{2R}{\rqu}\bigg\{\frac{\pi\kb T |t|}{\hbar} %
+ \ln(1-e^{-2\pi\kb T|t|/\hbar}) }\qquad\qquad\qquad\qquad\nonumber \\
& & \mbox{}+i \frac{\pi}{2}\text{sign}(t) - \ln (\pi/{\cal T})\bigg\}
\end{eqnarray}
where the constant term $\ln(\pi/{\cal T})$ will allow $P(E)$ to be
approximately normalized.  This is essentially the
result\cite{Wilhelm:2001} of Wilhelm, \etal, and is generally
appropriate for dealing with the high-frequency part of Eq.~(\ref{rezteq})
due to the relatively small values of \ztd\ and \ceff.  However, this
form may also be used for a sufficiently short and narrow section of
transmission line.

In the opposite limit, for which ${\cal T}< 1$, we find that the approximate
analytic result
\begin{equation}\label{ktapxeq}
k_{{\cal T}}(t,{\cal T}) \approx \pi e^{-|t|/\tau}\cot{\cal T}
\end{equation}
holds for times $|t|/\tau > {\cal T}$.  For typical temperature scales
available in a dilution refrigerator ($T=20$--\amount{400}{mK}), the
logarithmic term is very small when (\ref{ktapxeq}) is applicable. 
We can therefore neglect it, and find\cite{Lu:2002a} that
\begin{eqnarray}\label{kseq}
K_{s}(t)=-\frac{2R}{\rqu}\bigg\{\pi\kb T %
|t|/\hbar+\frac{\pi}{2}[\cot({\cal T})\nonumber\\
 \mbox{}-i \text{sign}(t)](e^{-|t|/\tau}-1)\bigg\}
\end{eqnarray}
This result is typically most useful for dealing with finite $RC$ lines,
for which $\tau^{-1}$ is typically on the order of
\amount{10^{7}}{s^{-1}}. 

Finally, for $1\alt{\cal T}\alt 10$, both the logarithmic term and
$k_{{\cal T}}$ are of the same magnitude for the relevant time scales.
The analytic forms for $K(t)$ in (\ref{kleq}) and (\ref{kseq}) must then
neglect some potentially important term.

\subsubsection{Calculation of $P(E)$}

Having obtained analytic forms for $K(t)$, a straightforward application
of Eq.~(\ref{peeq}) allows one to calculate $P(E)$. Letting $g=\rqu/R$,
we have from Eq.~(\ref{kleq}) for the large ${\cal T}$ limit
\begin{equation}\label{pleq}
P_{l}(E)=\frac{(\pi/{\cal T})^{2/g}}{2\pi^2\kb T}
\text{Re}\left[e^{-\pi/g}B\left(\frac{1}{g}-\frac{i E}{2\pi\kb T}, 1-\frac{2}{g}\right)\right]
\end{equation} 
where $B(x,y)$ is the beta function, in agreement with Wilhelm,
\etal\cite{Wilhelm:2001} For \ztd\ we have from (\ref{rezteq}) that
$R=\kappa_{2}\ztd$ and $\tau=\kappa_{2}\ceff\ztd$.  While(\ref{pleq}) is only
valid for $g>1$, in terms of \ztd\ this condition becomes
$\ztd<\rqu/\kappa_{2}$, so that the result remains valid for quite large
$\ztd$ when $\kappa_{2}$ is small.

In the small ${\cal T}$ limit, we use (\ref{kseq}) to obtain\cite{Lu:2002a}
\begin{eqnarray}\label{pseq}
\lefteqn{P_{s}(E)=\frac{\tau}{\pi\hbar}e^{\gamma_{3}({\cal T}, g)}%
\text{Re}\big[e^{-i\pi/g}\gamma_{2}({\cal T},g)^{-\gamma_{1}({\cal T},g)}}\nonumber\\
& & \times\left\{\Gamma\left(\gamma_{1}({\cal T},g)\right)%
-\Gamma\left(\gamma_{1}({\cal T},g), \gamma_{2}({\cal T},g)\right)\right\}\big]
\end{eqnarray}
where $\Gamma(x,y)$ is the incomplete gamma function, and
$\gamma_{1}=\frac{\pi}{g}{\cal T}-i\frac{E\tau}{\hbar}$,
$\gamma_{2}=\frac{\pi}{g}(\cot{\cal T}-i)$ and
$\gamma_{3}=\frac{\pi}{g}\cot{\cal T}$.  Typically, this will be applied
to a finite $RC$ line, for which $R=\kappa_{1}\rl\ell$ and
$\tau=\rl\cl\ell^2/\sqrt{6}$.  In some cases, for a short $RC$ line, it
may be more correct to use $P_{l}(E)$, and substitute the appropriate
forms for $R$ and $\tau$ in Eq.~(\ref{pleq}) instead.

We show $P_{s}(E)$ in  Fig.~\ref{zrcfig}(b)  for the same transmission
line parameters as used in Fig.~\ref{zrcfig}(a).  For comparison, we
also show the $T=0$ form for an infinite transmission line given by
$P_{\text{inf}}(E)=\sqrt{eV_0/2\pi E^{3}}e^{-eV_0/2E}$ where
$eV_{0}=(4\rl\kappa_{1}/\rqu)(e^2/2\cl)$.  Even at $T=\amount{10}{mK}$,
 $P_s(E)$ is significantly broader than $P_{\text{inf}}(E)$.  While it
is possible to obtain an analytic form for $P_{\text{inf}}(E)$ by
expanding Eq.~(\ref{koteq}) in the high temperature limit, the resulting
expression is of limited use for $E\neq 0$, since it involves only even
powers of $E$ and therefore cannot satisfy detailed
balance.\cite{Ingold:1992}  As a result, such an expression cannot be
used to calculate \iv\ characteristics, for instance, whereas $P_{s}(E)$
in (\ref{pseq}) can.

Ultimately, we are interested in calculating $P_{\text{tot}}(E)$ for the
total impedance \rezt\ seen by the tunneling electrons.  If we were to
calculate the total kernel $K(t)$ for the decomposition in
(\ref{rezteq}), it would in general include all the terms in
(\ref{kotaneq}). We were unable to find an analytic form for
$P_{\text{tot}}(E)$ under those circumstances. However, given the
decomposition (\ref{rezteq}), it is possible to write
$K(t)=K_{\text{lf}}(t)+K_{\text{hf}}(t)$ where $K_{\text{lf}}(t)$ and
$K_{\text{hf}}(t)$ correspond to the low- and high-frequency parts of
\rezt, with corresponding $P_{\text{lf}}(E)$ and $P_{\text{hf}}(E)$. 
The total $P(E)$ is then given by the convolution
$P(E)=P_{\text{lf}}(t)\ast
P_{\text{tot}}(E)=\int_{-\infty}^{\infty}P_{\text{lf}}(E-E')P_{\text{hf}
}(E')\,dE'$, which can be performed numerically.

\subsection{Multisection Transmission Lines}\label{multi}

In our particular case, the sample leads do not have a single width.
Instead, they broaden in sections from \amount{0.4}{\mu m} (section 1)
to \amount{375}{\mu m} (section 4) as detailed in Table~\ref{rctable}
below.  As a result, we must generalize (\ref{zrclfeq}) to allow for
the possibility of multiple sections.  In general, we use
Eq.~(\ref{lleq}) for a loaded transmission line, beginning closest to
the SET with section 1.  For this section, $Z_{L}$ is taken to be the
impedance of the second section, which is in turn terminated by the
following sections.  This cascading process is taken to end at our
macroscopic contact pads, which are so broad as to provide very little
impedance, and we therefore take $Z_{L}=0$ for the last section, so that its
impedance is given by (\ref{zrceq}).  We also ignore a short
($\ell=\amount{1}{\mu m}$) section with $w=\amount{100}{nm}$ since it
contributes only \amount{50}{\Omega} to $\zl(0)$, and its associated
$P(E)$ is very sharply peaked around $E=0$.

\begin{figure}
\includegraphics[width=3.1in]{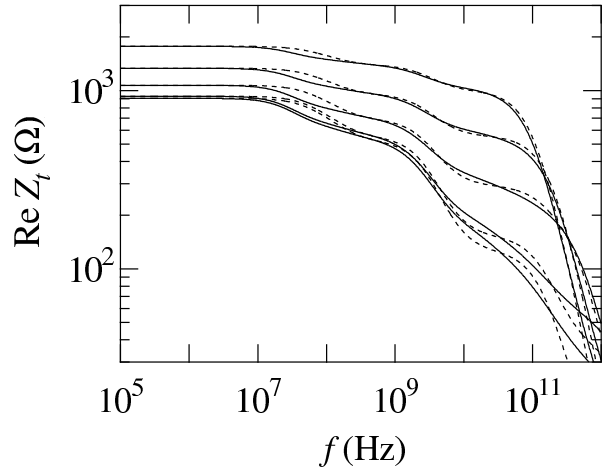}
\caption{\label{frcline} \rezt\ for four cascaded $RC$ lines and a
ground plane, using the parameters given in Table~\ref{rctable}, for
(top to bottom) $\ztd=6445$, 3227, 1291, 258 and \amount{100}{\Omega}. 
Solid lines: exact calculation of \rezt.  Dashed lines: approximate
version of \rezt\ as described in the text.}
\end{figure}

If we were to use the exact form for \zl\ given by the above cascading
procedure, it would be too complex to be of  use.  Fortunately, a simple
approximation gives fairly accurate results.  We take
\begin{equation}\label{frceq}
\text{Re}[\zl(\omega)] = \sum_{i}\frac{\rli\ell_{i}}{1 +%
 (\omega\rli\cli\ell_{i}^{2}/\sqrt{6})^2}
\end{equation}
where \rli\ and \cli\ are the resistance and capacitance per unit length
of section $i$, and $\ell_{i}$ is its length.  We use the width $w_{i}$
and length $\ell_{i}$ of each section along with the 2DEG sheet
resistance $\rsq=\amount{20}{\Omega}$ and depth $h=\amount{50}{nm}$ to
calculate\cite{Collin:1992} $\rli\approx\rsq/(w_{i}+5.8h)$ and
$\cli\approx\varepsilon\varepsilon_{0}(w_{i}/h+1.393)$, where
$\varepsilon=13$ is the dielectric constant of GaAs.  To find an
approximate form for \rezt\ we use the result (\ref{frceq}) for \zl\ in
the decomposition (\ref{rezteq}). For comparison, we plot both this
approximate result as well as the  exact one obtained from the full form
for \rezt\ in (\ref{zteq})--(\ref{yteq}) and repeated applications of
(\ref{lleq}) versus frequency in Fig.~\ref{frcline} for different values
of \ztd.  The agreement is very good, especially considering the number
of approximations required to develop a tractable approximate form for
\rezt. The approximate version tracks the exact result very well except
between the various corner frequencies of the transmission line
sections, where its slope is generally too small. Agreement is better
overall for larger values of \ztd, but even for the smallest values is
still acceptable.
\begin{table}
\caption{\label{rctable}Transmission line parameters for the various
sections of the sample leads, with $w$ and $\ell$ in $\mu$m, \rl\ in
M$\Omega$/m, \cl\ in nF/m, $\rl\ell$ in $\Omega$, and ${\cal T}$
calculated for $T=\amount{100}{mK}.$}
\begin{ruledtabular}
\begin{tabular}{c|ddddcc}
section & $w$ & \ell & \rl & \cl & $\rl\ell$ & ${\cal T}$ \\ \hline
1 & 0.4 & 9 & 29 & 1.08 & 260 & 37 \\
2 & 1 & 57 & 15.5 & 2.46 & 884 & 0.76 \\
3 & 10 & 253 & 1.9 & 23 & 491 & $3.2\e{-2}$ \\
4 & 20 & 375 & 1.0 & 46 & 375 & $1.5\e{-2}$ \\
\end{tabular}
\end{ruledtabular}
\end{table}

We can calculate $P_{\ell}^{(j)}(E)$ for tunneling through junction $j$
for the four section transmission line by choosing either $P_{l}(E)$ or
$P_{s}(E)$ for a given section based on its value of ${\cal T}$ in
Table~\ref{rctable}, and numerically convolving the four functions
through
\begin{equation}
P_{\ell}^{(j)}(E)=P_{1l}^{(j)}(E)\ast P_{2s}^{(j)}(E)\ast P_{3s}^{(j)}(E)%
 \ast P_{4s}^{(j)}(E).
\end{equation}
While somewhat time consuming, this procedure needs to be performed only
once for a given temperature since the 2DEG beneath the transmission
lines is not affected by the Au gates, and so the transmission line
parameters do not change with \vg.

\begin{figure}
\includegraphics[width=3.0in]{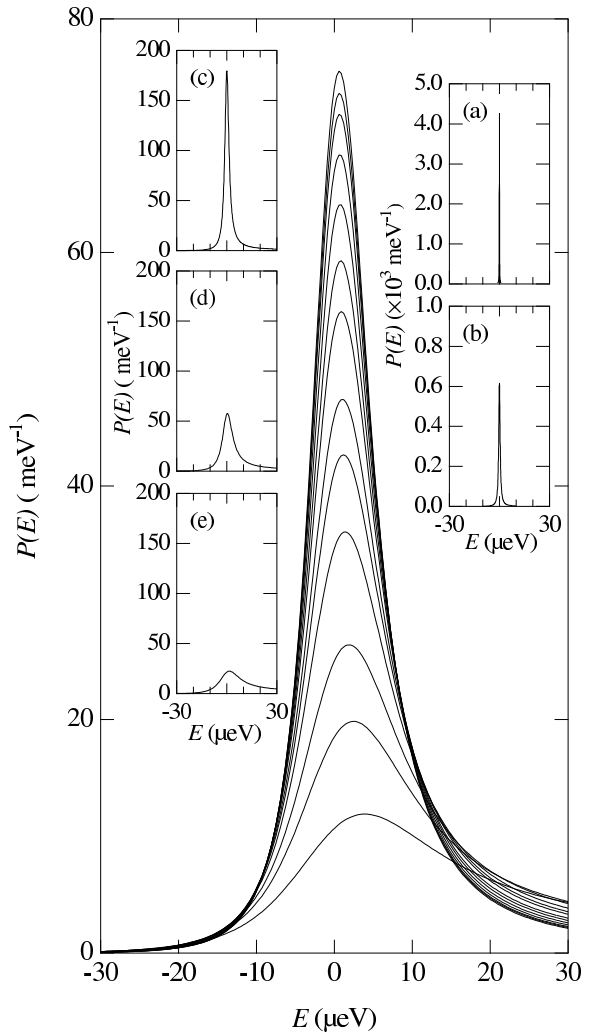}
\caption{\label{pefig}Calculated $P_{\text{tot}}^{(1)}(E)$ for S2,
based on the transmission line parameters from Table~\ref{rctable} for a
series of values of \ztd.  Top to bottom: $\ztd=0$, 65, 129, 258, 430,
645, 860, 1291, 1613, 2151, 3227, 4302 and \amount{6445}{\Omega}. 
Insets: $P_{2D}^{(1)}(E)$ for \ztd\ equal to (a) 65 (b) 430 (c) 1291  (d)
3227 and (e) \amount{6445}{\Omega}.  Note the scale change for $\ztd=65$
and \amount{430}{\Omega}.}
\end{figure}

Finally, we then calculate the total $P_{\text{tot}}^{(j)}(E)$ for
tunneling through junction $j$ by convolving $P_{\ell}^{(j)}(E)$ with
$P_{2D}^{(j)}(E)$ calculated from $P_{l}(E)$ for the appropriate value
of \ztd.  This procedure typically must be performed  many times, but
can be done relatively quickly.  Results for $P_{\text{tot}}^{(1)}(E)$
for tunneling through junction 1 of S2 are shown in
Fig.~\ref{pefig} for a series of different values of \ztd.  For
$\ztd=0$, we take $P_{\text{tot}}^{(1)}(E)=P_{\ell}(E)$, which is
already relatively broad, with a width several microvolts.  In contrast,
for small \ztd, $P_{2D}^{(1)}(E)$ is very sharply peaked around $E=0$
and approximates a delta function, as can be seen in the insets (a) and
(b) in Fig.~\ref{pefig}.  As a result, $P_{\text{tot}}^{(1)}(E)$ is not
strongly affected by $P_{2D}^{(1)}(E)$ until
$\ztd\agt\amount{200}{\Omega}$.  Finally, for sufficiently large \ztd,
$P_{2D}^{(1)}(E)$ begins to dominate and $P_{\text{tot}}^{(1)}(E)$
becomes very broad, indicating the high probability of inelastic
transitions.  Overall, the trend is for the transmission line to
dominate energy exchange for small \ztd, while the 2DEG dominates energy
exchange for large \ztd.

\subsection{Calculation of \iv\ Curves}

To calculate the \iv\ characteristics for the S-SET, we use a master
equation approach\cite{Schon:1998a} in which we assume that only two
charge states, $N$ and $N+1$ where $N$ is the number of Cooper pairs,
are important.   This should be a valid approach for temperatures and
biases small compared to the charging energy \ec\ of the S-SET\@.  We
begin by calculating the free energy change for changing the island
charge from $N$ to $N+1$ (or vice versa) due to tunneling through
junction $j$.  We find
\begin{eqnarray}
\delta f^{(j)}_{N\rightarrow N+1} & = & -\delta f^{(j)}_{N+1\rightarrow N}\nonumber \\
& = & 4\ec(2N -n_{g}+1) -(-1)^{j}2\alpha_{j}eV
\end{eqnarray}
where $n_{g}=\vg\cg/e$ is the gate charge and
$\alpha_{j}=\textstyle{\frac{1}{2}}+(-1)^{j}(\co-\ct)/2\csig$ is the
fraction of the bias voltage $V$ appearing across junction $j$.  We then
use Eq.~\ref{gameq} to find the tunneling rates in terms of the sample
parameters.

The master equation can be solved exactly when only two charge states
are considered.\cite{Schon:1998a}  Doing so, and using the detailed
balance relation $P(-E)=e^{-E/\kb T}P(E)$, we find
\begin{equation}\label{iveq}
I(V) = \frac{\pi^2\Delta^2}{32 e \rk} \frac{\sinh\left(\frac{eV}{\kb%
 T}\right)}{\frac{\tilde{r}_{2}^{2}}{\widetilde{P}_{2}}%
\cosh\left(\frac{\delta f^{(1)}}{2 \kb T}\right)%
+\frac{\tilde{r}_{1}^{2}}{\widetilde{P}_{1}}%
\cosh\left(\frac{\delta f^{(2)}}{2 \kb T}\right)}%
\end{equation}
where $\delta f^{(j)}$ is the change in free energy for tunneling in the
electrostatically favorable direction ($N+1\rightarrow N$ for junction 1
and $N\rightarrow N+1$ junction 2), $\tilde{r}_{j}=R_{j}/\rk$,
$\widetilde{P}_{j} = (P_{j}^{+}P_{j}^{-})^{1/2}$,
$P_{j}^{\pm}=P_{\text{tot}}^{(j)}(\mp\delta f^{(j)})$, and $\rk=h/e^2$
is the resistance quantum.

\section{Experimental Results}

\subsection{Measurements}

We have performed electrical measurements on the two samples described
in Table~\ref{params} in a dilution refrigerator at mixing chamber
temperatures ranging from $T=20$ to \amount{400}{mK}.  Measurements were
performed in a four-probe voltage biased configuration, with the bias
applied symmetrically with respect to ground, in a shielded room using
battery powered amplifiers.  High frequency noise was excluded with
$\pi$-filters at room temperature and microwave filters at the mixing
chamber.

Because the S-SET and 2DEG are electrically isolated from each other at
dc, we were able to measure their conductances \gset\ and \gtd\
separately.  In both cases the conductance was measured by applying an
ac  bias voltage at \amount{11}{Hz} and measuring the resulting current
using standard lock-in techniques. The bias voltage used was 3 and
\amount{5}{\mu V} respectively for the S-SET and 2DEG\@.  We also
performed measurements of dc \iv\ characteristics of the S-SET\@.  For
the 2DEG, the current contacts were positioned on opposite sides of the
two QPCs, so that \gtd\ measured the series combination of their
conductances.  Since the S-SET sees the QPCs in parallel, we take
$\ztd=1/(4\gtd)$ for the pool geometry.  For the stripe geometry we
cannot measure the stripe resistance \rstr\ or capacitance \cstr\
directly, although we can estimate them from the sample design.  We
expect that in this geometry $\ztd\approx\rqpc/2$ at low frequencies and
$\ztd\approx\rstr/2\approx\amount{100}{\Omega}$ at high frequencies, as
discussed above.  While the \iv\ measurements of the S-SET were usually
performed at a fixed gate voltage \vg, the conductance measurements were
typically performed versus \vg\ for a variety of temperatures.  In the
pool geometry all six Au gates were tied together and \vg\ swept from 0
to \amount{-1}{V}, while in the stripe configuration only the four
outermost gates were swept.

\begin{figure}
\includegraphics[width=3.0in]{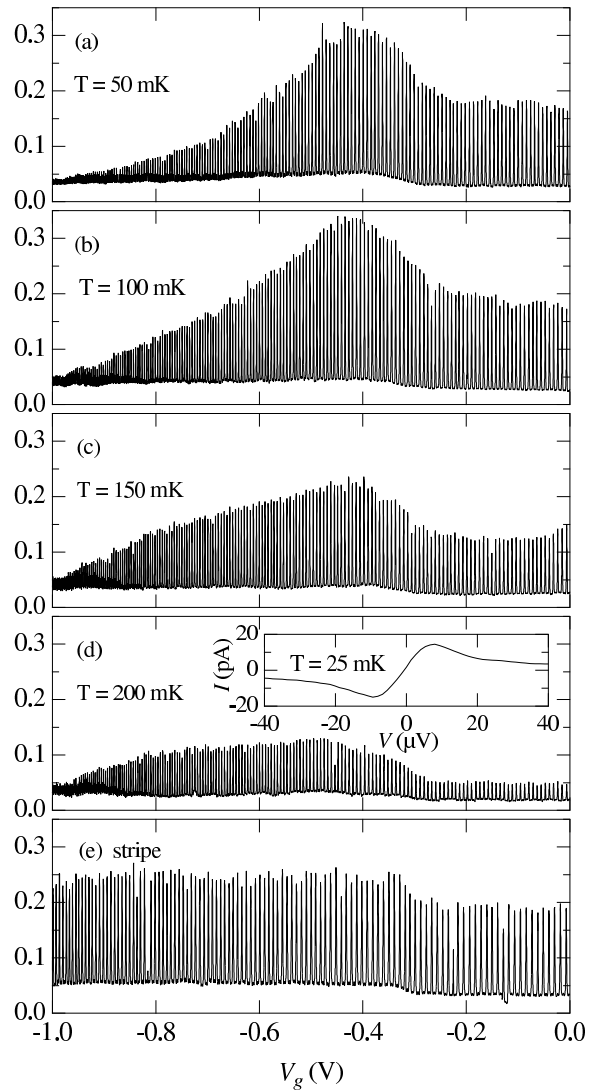}
\caption{\label{gvsvgfig} \gset\ vs.\ \vg\ for S2 in the pool geometry
for $T=$ (a) 50, (b) 100, (c) 150, and (d) \amount{200}{mK}.  In (e) we
show \gset\ vs.\ \vg\ in the stripe geometry for $T=\amount{25}{mK}$. 
The inset to (d) shows an \iv\ characteristic for S2 at
$T=\amount{25}{mK}$ with the 2DEG left unconfined
($\vg=\amount{0}{V}$).}
\end{figure}

Results of these measurements are shown for S2 in Fig.~\ref{gvsvgfig}
for mixing chamber temperatures ranging from 50 to \amount{200}{mK}.  As
\vg\ is made more negative, the S-SET conductance oscillates due to the
effects of the Coulomb blockade; the oscillations are clearly visible in
Fig.~\ref{gvsvgfig}(a)--(e).  We show results for the pool geometry for
various temperatures in Fig.~\ref{gvsvgfig}(a)--(d), and results for the
stripe geometry in Fig.~\ref{gvsvgfig}(e).  In both geometries, the
envelope of the oscillations is relatively flat until
$\vg\approx\amount{-0.3}{V}$, at which point the amplitude of
oscillations begins to increase.  For the pool geometry, the envelope
continues to rise until $\vg\alt\amount{-0.4}{V}$, at which point it
begins to fall again.  For the lower temperatures
($T\leq\amount{100}{mK}$), the drop with more negative \vg\ is quite steep,
whereas for the higher temperatures
($\amount{100}{mK}<T<\amount{200}{mK}$) the drop is more gradual.  For
less negative \vg, the envelope generally rises as $T$ decreases, and
tends to saturate below $T\approx\amount{100}{mK}$.  For more negative
\vg\ (beyond the maximum in the envelope at $\vg=\amount{-0.41}{V}$) the
envelope rises as $T$ decreases until $T\approx\amount{100}{mK}$, below
which it decreases.

Since \gtd\ decreases as \vg\ becomes more negative, the above behavior
indicates that \gset\ varies non-monotonically as \gtd\ decreases, first
rising and then falling.  A decrease in \gset\ as \gtd\ decreases is
expected.  Physically, a larger \gtd\ tends to damp phase fluctuations,
promoting superconducting behavior and therefore a higher \gset. 
Alternatively, we can say that for low energies, the probability $P(E)$
of exchanging energy with the environment increases as \gtd\ increases,
as can be seen in Fig.~\ref{pefig}.  A higher \gtd\ therefore implies a
higher probability of elastic (or nearly elastic) transitions, and a
higher \gset\ at low bias.  In contrast, the decrease in \gset\ for
$\vg\agt\amount{-0.4}{V}$ does not fit in with this general picture of
energy exchange with the environment.  While nonmonotonic behavior with
\gtd\ can be expected,\cite{Wilhelm:2001} it is generally associated
with a crossover from the non-linear to linear portions of the \iv\
characteristic.  We find that this is true in our simulations as well.

In contrast, for our experiments \gset\ was always measured in the linear
part of the \iv\ characteristic.  We show in the inset to
Fig.~\ref{gvsvgfig}(d) the \iv\ characteristic for S2 at
$T=\amount{20}{mK}$ for $\vg=\amount{0}{V}$, \ie, when the 2DEG is
unconfined.  The \iv\ characteristic is clearly linear to \amount{\pm
8}{\mu V}, so that our \amount{3}{\mu V} rms bias should be firmly in
the linear regime.  The non-monotonic behavior we observe cannot then be
associated with changes in $P(E)$, and must arise from other physics.

We can find a clue as to the source of this behavior by examining \gset\
vs.\ \vg\ for the stripe geometry as shown in Fig.~\ref{gvsvgfig}(e). In
this geometry the envelope also begins to rise at
$\vg\approx\amount{-0.3}{V}$.  However, the rise is weaker, with the
envelope increasing by only roughly half what is observed in the pool
geometry.  Furthermore, there is no decline in \gset\ as \vg\ is made
more negative.   This last observation is in agreement with our model
for \ztd\ in this geometry, which predicts that \ztd\ approaches
$\rstr/2$ at the higher frequencies which dominate our measurement of
\gset.  As a result, in this case $P_{\text{tot}}(E)$ never broadens for
more negative \vg\ as it does in the pool geometry, so that no decrease
in \gset\ is observed.  Once again, the environmental theory cannot
explain the reduction in \gset\ for less negative \vg.

One possible explanation for this decrease  is that there are changes in
the offset charge of the S-SET island that occur on a time scale
 short in comparison to the time constant of the lock-in, but long in
comparison to the time scales associated with environmental
fluctuations.  It is well known that charge fluctuations in the
substrate give rise to $1/f$ noise in SET-based
electrometers.\cite{Visscher:1995,Krupenin:1998} Such charge noise
typically has a magnitude of $(S_{Q})^{1/2}\sim10^{-4}$--$10^{-3}\ehz$
at \amount{10}{Hz} and a cutoff frequency (above which the intrinsic SET
noise dominates) of about 100--\amount{1000}{Hz}.  Let us assume that in
our case the $1/f$ noise is somewhat larger than is typical, say
$(S_{Q})^{1/2}\sim 4\e{-3}\ehz$ at \amount{10}{Hz}, due to the presence
of the 2DEG\@.  If we write $S_{Q}=1.6\e{-4}e^2/f\equiv S_{0}/f$, then
the expected mean square charge variance\cite{Kogan:1996} between
frequencies $f_{1}$ and $f_{2}$ is given by
$\langle\sch^2\rangle=S_{0}\ln(f_{2}/f_{1})$. Taking
$f_{1}=\amount{0.1}{Hz}$ and $f_{2}=\amount{1000}{Hz}$ we find
$\langle\sch^2\rangle^{1/2}\sim 4\e{-2}e$, so that a typical variance of
a few hundredths of an electronic charge is not unreasonable.

In a voltage biased configuration such as ours, these fluctuations would
have the effect of averaging the measured current over an ensemble of
charge states centered around the gate charge $n_{g}$. Similar effects
have been seen in measurements of other S-SET systems.\cite{Eiles:1994}
Since the S-SET current is sharply peaked around the charge degeneracy
points, we expect that any such charge averaging would tend to reduce
the measured peak current, and therefore the conductance \gset.  It is
therefore possible that the reduction in \gset\ for less negative \vg\
arises due to increased charge averaging as the electrons in the 2DEG
become less confined.  We examine this possibility in more detail in the
following section.

\subsection{Comparison with Theory}

\begin{figure}
\includegraphics[width=3.0in]{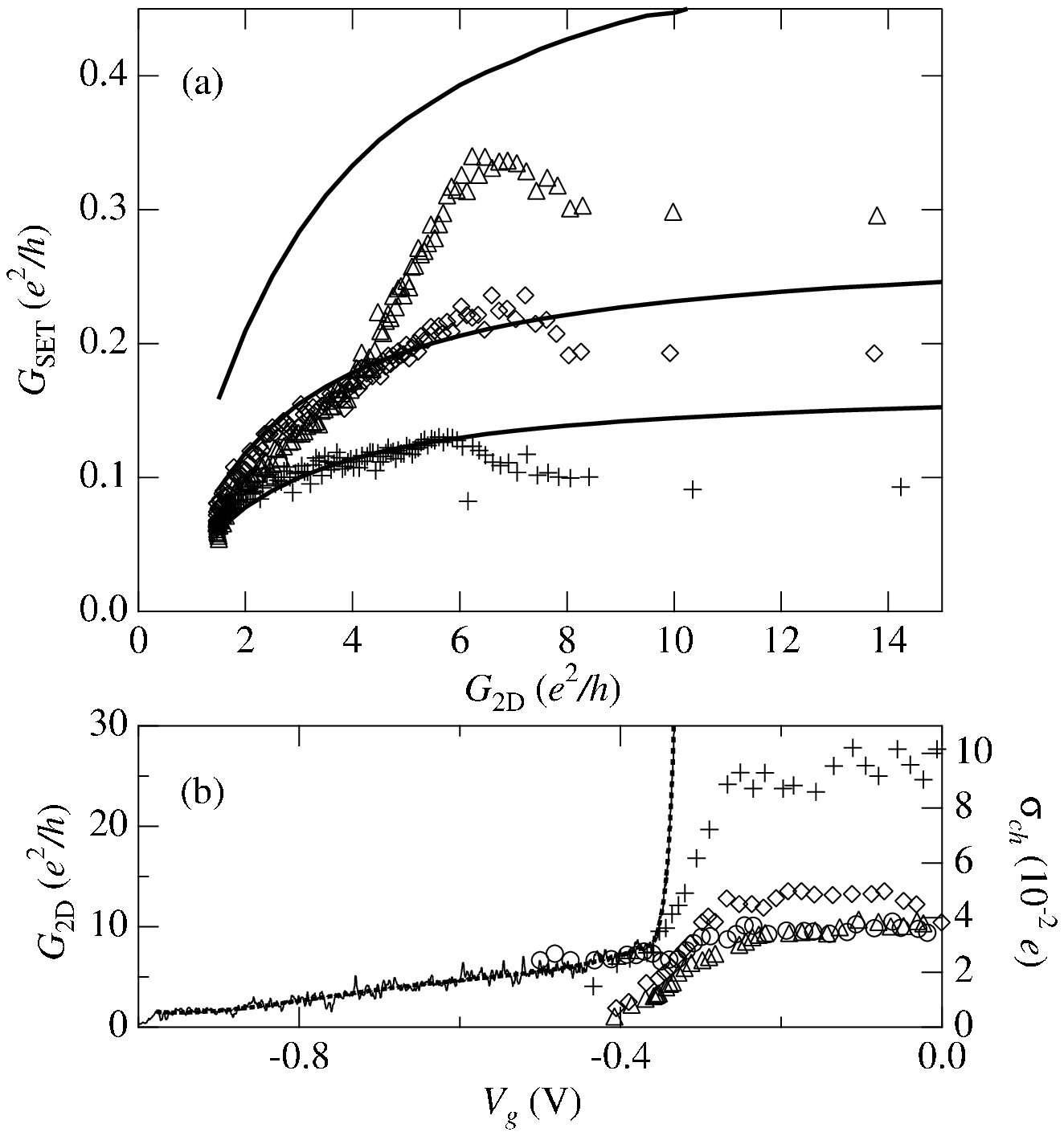}
\caption{\label{gvsgdfig} (a) Peak values of \gset\ versus \gtd for
$T=\amount{100}{mK}$ ($\triangle$), \amount{150}{mK} ($\Diamond$), and
\amount{200}{mK} ($+$).  Calculated values \gsc\ scaled to equal \gset\
at its maximum value at $\gtd^{\text{max}}$ at \amount{200}{mK} are
shown as the heavy solid lines. (b) On the left axis we plot measured
\gtd\ versus \vg\ (solid line), and the smooth function fitted to \gtd\
vs.\ \vg\ (heavy dashed line).  On the right axis we plot values of
\sch\ in the pool geometry for $T=\amount{100}{mK}$ ($\triangle$),
\amount{150}{mK} ($\Diamond$), and \amount{200}{mK} ($+$).  We also show
\sch\ for the stripe geometry
($\circ$).   }
\end{figure}

To compare our measurements with theory, we must plot the measured
\gset\ versus \gtd.  Having measured \gtd\ versus \vg, we can fit a
smooth function to the measured \gtd\ and use it to convert \vg\ to an
approximate \gtd.  In Fig.~\ref{gvsgdfig}(b) below we show both the
measured \gtd\ and the fitted function versus \vg\ in units of conductance quanta
$G_{0}=e^2/h$.  While the two are nearly indistinguishable in
the figure, this procedure in only useful for a limited voltage range. 
For $\vg\agt\amount{-0.31}{V}$ ($\gtd\agt 200G_{0}$) the measured values
of \gtd\ become unreliable.

Having converted \vg\ to \gtd\, we must now plot the envelope of \gset\
versus \gtd; we do so by recording the positions of the peaks in \gset\
versus \vg\ and using our fitting function to convert to \gtd.  The
results of this process are shown for $T=100$, 150 and \amount{200}{mK}
in Fig.~\ref{gvsgdfig}(a).  As can be seen in Fig.~\ref{gvsvgfig}, our
data show no sign of $2e$
periodicity,\cite{Tuominen:1992,Eiles:1993,Amar:1994,Joyez:1994} even at
the lowest temperatures.  A similar lack of $2e$ periodicity was
observed by the Berkeley group as well.\cite{Kycia:2001}  At the present
time, it is unknown whether this is an intrinsic feature of S-SET/2DEG systems
or is due to the inherent difficulties of excluding high-frequency noise
from all possible sources, including the substrate.\cite{Covington:2000}
In any case, we presume that since our data are strictly $e$-periodic
that probabilities of finding the S-SET island with either an even or
odd number of electrons are approximately equal.  This will have little
overall effect on our analysis, since at any given value of $n_{g}$ only
one charge state contributes significantly to the current, the other
being between charge degeneracy points and therefore in a low-current
regime. Overall, we would then expect an additional reduction of a
factor of roughly two in the measured current above and beyond that due
to any other reason.

In order to compare our results with theory, we calculate
$P_{\text{tot}}(E)$ according to the procedure described in
Section~\ref{multi} above.  We then use (\ref{iveq}) above to obtain a
calculated conductance \gsc.  In general, the conductance calculated
directly from (\ref{iveq}) overestimates the measured conductance
\gset\ by a significant factor (on the order of 40--50).  Such
discrepancies are not uncommon in small tunnel junction
systems,\cite{Tuominen:1992,Eiles:1994,Kycia:2001} but nevertheless complicate
comparison with theory.  The best we can achieve is to compare the
relative change in \gset\ and \gsc, but to do so requires that we pick
some point of reference at which we will scale \gsc\ to make it equal to
\gset.

\gset\ reaches its maximum value at some value of \gtd\ which for all
the temperatures considered here is approximately $6.5 G_{0}$; we call
this value $\gtd^{\text{max}}$, which may vary slightly with
temperature.  For $\gtd>\gtd^{\text{max}}$, \gset\ decreases, behavior
which cannot be accounted for through changes in the environment,
making this region unsuitable for choosing a reference point.  For 
$\gtd<\gtd^{\text{max}}$, we assume for simplicity that there is no
charge averaging, and choose the point at which \gset\ reaches its
maximum at $T=\amount{200}{mK}$ as our reference point.

We plot \gsc\ versus \gtd, choosing our reference point as discussed
above, in Fig.~\ref{gvsgdfig}(a) as the heavy solid lines.  All curves
are scaled by the same factor ($\approx44$) so that we can get some
sense of the agreement in terms of temperature dependence as well as
dependence on \ztd.  Generally speaking, the agreement at $T=150$ and
\amount{200}{mK} is quite good for $\gtd<\gtd^{\text{max}}$; both
calculated curves for \gsc\ track the experimental values of \gset\
nearly exactly. We note that to obtain this level of agreement, we have
used only one variable parameter, namely the scaling factor.  All other
parameters used in the theory are derived from experimentally measured
quantities.  Given the number of approximations involved, the level of
agreement for these two temperatures is quite remarkable.

For temperatures higher than $T=\amount{200}{mK}$, the S-SET conductance
begins to rise again.  However, we expect the environmental theory
discussed here to be applicable only for temperatures satisfying the
Coulomb blockade condition\cite{Grabert:1998} $\kb T \ll
\ec\rqu/(2\pi^2\text{Re}[Z_{t}(0))]$, which for S2 corresponds to
$T\approx\amount{240}{mK}$ for typical values of $\text{Re}[Z_{t}(0)]$.
Failure of the theory between $T=200$ and \amount{250}{mK} is in good
agreement with this condition.

At $T=\amount{100}{mK}$ and below, the theory also disagrees with the
experimental results.  Specifically, the measured \gset\ does not rise
as rapidly with decreasing $T$ as predicted by theory.  Furthermore, the
dependence of \gset\ on \gtd\ changes from sublinear to superlinear, so
that the S-SET conductance depends more strongly on the environmental
impedance than theory predicts.  This trend is accentuated at lower
temperatures, as can be seen in Fig.~\ref{gvsvgfig}(a) for
$T=\amount{50}{mK}$.  In this case \gset\ is an even stronger function
of \gtd, and \gset\ is generally speaking slightly smaller that at
$T=\amount{100}{mK}$ (not larger as would be expected from the theory). 
One possible explanation for a saturation of \gset\ would be that the
electron temperature stops decreasing for some temperature below
\amount{100}{mK}.  While it is likely that our electron temperature
saturates (data at \amount{20}{mK} differs only very slightly from the
\amount{50}{mK} data), such effects would not explain the change in
dependence on \gtd, or a decrease in \gset\ from 100 to \amount{50}{mK}.

There is also a lower temperature bound for applicability of the
environmental theory, set by the condition
$P_{\text{max}}^{(i)}E_{J_{i}}\ll 1$ where $P_{\text{max}}^{(i)}$ is the
maximum value of $P_{\text{tot}}^{(i)}(E)$.  This condition must be
satisfied for the perturbative result for the tunneling rate in
Eq.~(\ref{gameq}) to be satisfied.  In our case, we find that for
tunneling through junction 1 (for which both $E_{J}$ and
$P_{\text{max}}$ are larger), $P_{\text{max}}^{(1)}\approx 52$, 41 and
$\amount{34}{meV}^{-1}$, all at $\gtd=6.5G_{0}$ and at $T=100$, 150 and
\amount{200}{mK}, respectively.   In that case we find that
$P_{\text{max}}^{(1)}E_{J_{1}}=1.4$, 1.1, and 0.92 for the same
temperatures. In none of the cases is the condition for agreement with
Eq.~(\ref{gameq}) clearly satisfied, so that the agreement at 150 and
\amount{200}{mK} is perhaps better that might be expected.  Still, if it
were a failure of the perturbative expansion which is leading to the
disagreement at \amount{100}{mK}, we would expect the theory to agree
for sufficiently low \gtd\ that $P_{\text{max}}$ drops to its largest
value at \amount{150}{mK}.  In our case, this occurs at $\gtd\approx
4.0G_{0}$ so that based on this argument we would expect theory and
experiment to agree at \amount{100}{mK} over much of the range shown,
and only deviate for $4.0 G_{0}<\gtd<6.5G_{0}$.  Clearly, this
expectation does not hold for our data.

We now turn our attention to the range $\gtd>\gtd^{\text{max}}$, for
which \gset\ decreases, in contradiction to the environmental theory. As
discussed above, we consider the possibility that some form of charge
averaging, which increases as \gtd\ increases and the electrons in the
2DEG become more mobile, causes \gset\ to decline.  To test the
plausibility of this hypothesis, we calculate the average conductance
$\langle\gsc\rangle$ given by \begin{equation}\label{gscaveeq}
\langle\gsc\rangle=\int_{-\infty}^{\infty}w(n')\gsc(V,n')dn'
\end{equation} where $\gsc(V,n)$  is the calculated SET conductance
(including the scaling factor) calculated at bias $V$ and gate charge
$n$ and $w(n)=
\frac{1}{\sqrt{2\pi}\sch}\exp\left[-\frac{(n-n_{g})^2}{2\sch^2}\right]$
is assumed to be the probability of finding the SET in charge state $n$
when the gate charge is $n_{g}$.  We vary \sch\ to cause the
$\langle\gsc\rangle$ to exactly match the measured conductance at a
given \vg\, and plot the results in Fig.~\ref{gvsgdfig}(b) for $T=100$,
150, and \amount{250}{mK} in the pool geometry, and a mixing chamber
temperature of \amount{20}{mK} for the stripe geometry (estimated
electron temperature roughly 50--\amount{70}{mK}).  By construction,
$\sch=0$ for $\gtd < \gtd^{\text{max}}$ in the pool geometry.  For the
stripe geometry, $\sch$ saturates at about $2\e{-2}e$ for large negative
\vg, since in this geometry \gset\ never reaches as large a value as it
does in the pool geometry.  This result is physically reasonable, since
the electrons in the stripe are never as confined as they are in the
pool.  As \vg\ becomes less negative, \sch\ rises until it saturates (in
all cases) at $\vg\approx\amount{-0.27}{V}$, the voltage at which the
2DEG begins to deplete.  The saturation value for \sch\ is generally
reasonable, being about $4\e{-2}e$ for the stripe and for the the pool
at \amount{100}{mK}, and $5\e{-2}e$ and $9\e{-2}e$ at 150 and
\amount{200}{mK}.  While the cause of the increase in \sch\ at
\amount{200}{mK} is unclear, there does not appear to be significant
further rise in \sch\ for higher temperatures.

\begin{figure} \includegraphics[width=3.0in]{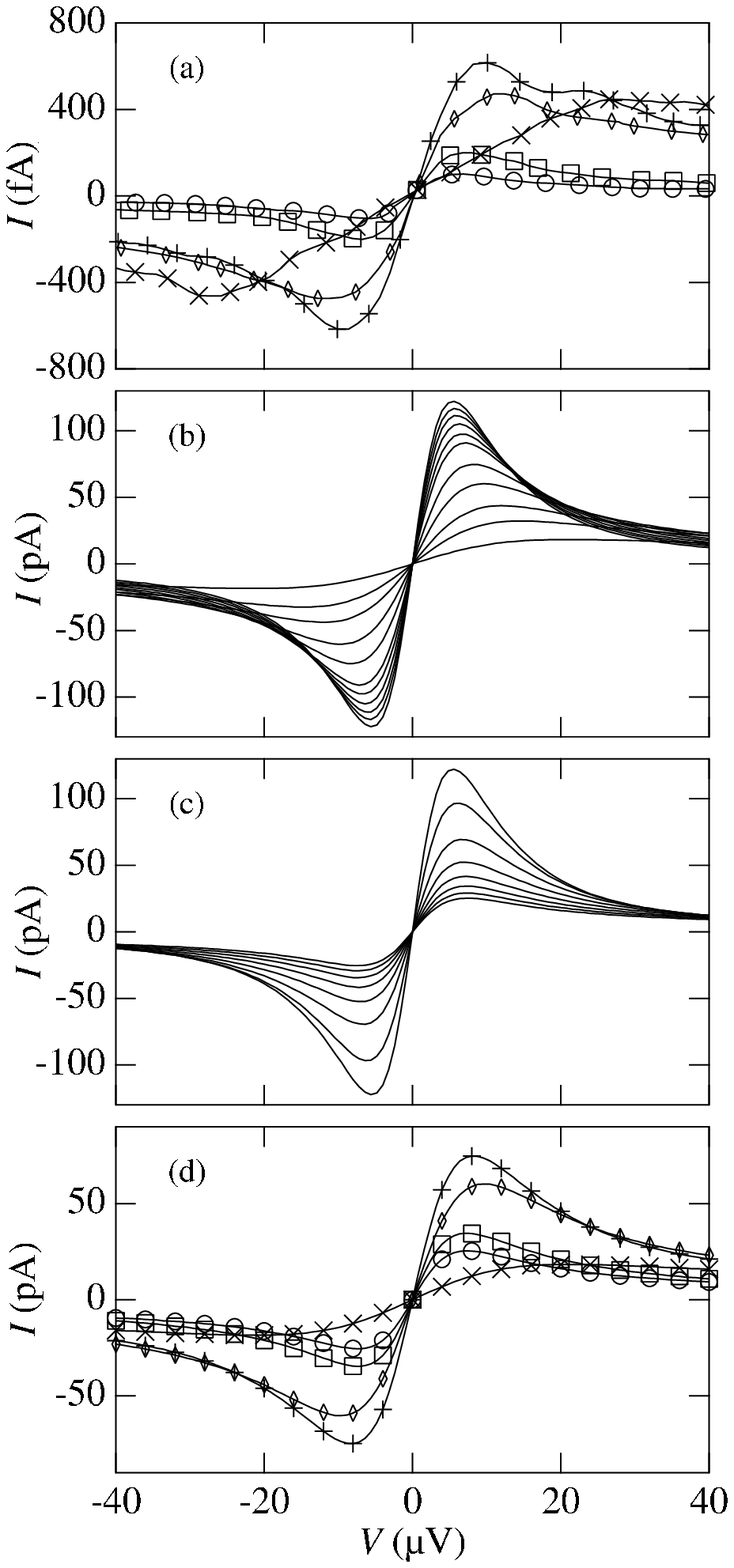}
\caption{\label{ivfig}(a) \iv\ characteristics for S1 at an estimated
electron temperature of \amount{100}{mK} for an unconfined 2DEG
($\circ$),  $\vg = \amount{-0.3}{V}$ ($\Box$), and $\ztd = 1613$ (+),
2151 ($\diamond$) and \amount{6453}{\Omega} ($\times$). (b) Calculated
\iv\ characteristics (top to bottom at peak) for $\ztd=0$, 129, 258,
430, 645, 860, 1613, 2151, 3227, 4302 and \amount{6453}{\Omega}.  Here
$\sch=0$ for all curves. (c) Calculated \iv\ characteristics for (top to
bottom) $\sch=0$, 1, 2, 3, 4, 5, 6 and $7\e{-2}e$.  Here $\ztd=0$ for
all curves. (d) Calculated \iv\ characteristics for an unconfined 2DEG
($\circ$),  $\vg = \amount{-0.3}{V}$ ($\Box$), and $\ztd = 1613$ (+),
2151 ($\diamond$) and \amount{6453}{\Omega} ($\times$). To fit the data
at $\vg = 0$ and \amount{-0.3}{V}, we use $\sigma_{ch} = 0.07$ and $0.05
e$, respectively.  For the remaining curves we take $\sigma_{ch} = 0$. }
\end{figure}

To provide further support for this idea, we examine \iv\
characteristics for S1 as shown in Fig.~\ref{ivfig}(a), which show the
evolution of the \iv\ characteristics when the 2DEG is increasingly
confined.  As the confinement is increased, the current initially rises
($\vg=\amount{-0.3}{V}$) at all voltages, while the peak current remains
at a fixed voltage.  For $\ztd=\amount{1613}{\Omega}$, the current has
increased again, but the peak current has begun to move to higher bias. 
For $\ztd=\amount{2151}{\Omega}$, the peak current has decreased and
moved again to yet higher bias, while the current at higher voltages has
generally begun to rise.  Finally for $\ztd=\amount{6453}{\Omega}$ the
\iv\ characteristic has become quite broad and the peak has moved
outward yet again. 

We can understand this evolution by examining the effects of variations
in \ztd\ and \sch\  on the \iv\ characteristics separately, as shown in
Fig.~\ref{ivfig}(b) and (c) respectively.  When \ztd\ alone is increased,
the peak current drops and the voltage at which the peak occurs
increases; at the same time, current at higher biases increases.  This
reflects broadening of $P_{\text{tot}}(E)$ as \ztd\ is increased, and a
higher probability of inelastic processes.  In contrast, when \sch\
alone is increased, the current decreases at all bias voltages, and the
voltage at which the maximum current appears is more or less fixed.

In order to obtain good agreement with experiment, we must include both
variations in \sch\ and \ztd, as shown in Fig.~\ref{ivfig}(d).  Here,
for $\vg=0$ and \amount{-0.3}{V}, we take $\ztd=0$ and vary \sch, while
for $\ztd=1613$, 2151, and \amount{6453}{\Omega} we take $\sch=0$. 
Overall the theory agrees with the experimental results quite well
(apart from an overall scaling factor), reproducing the initial rise in
current with no shift in peak current position, followed by a reduction in
current and an outward shift in peak position.  The agreement is poor
only for $\ztd=\amount{6453}{\Omega}$, for which the experimental
current is too large in relative terms.  Even here, however, the shape
of the \iv\ curve is reproduced nicely.  Additional data (not shown)
indicates that the S-SET current does usually drop sharply for large
\ztd, so that the amplitude of this particular \iv\ curve is probably
anomalously high.

\subsection{Discussion}

Overall, the agreement between our experimental results and calculations
seems quite good, particularly for $T=150$ and \amount{200}{mK}.  The
good agreement of \gsc\ and \gset\ for those temperatures, combined with
the accurate predictions of our model for the evolution of the \iv\
characteristics gives us confidence that our model, despite its
complexity, accurately describes our experimental system.  In
particular, it is clear that both the lead impedance and any impedance
which is coupled directly to the S-SET island must be included to give
accurate results.  In our particular case, charge averaging appears to
play an important role when the confinement of the 2DEG is reduced,
either for less negative \vg\ or when the stripe geometry is used. 
Overall, this improved understanding indicates that S-SET/2DEG systems
can be used to test the accuracy of the standard environmental theory in
a way which was not previously possible.

Generally speaking, the model of the environment presented in
(\ref{rezteq}) should be applicable to any system consisting of an
S-SET and its leads fabricated above a 2DEG, including that of the
Berkeley group.  The model also gives a simple explanation for the lack
of agreement between the experimental results of the Berkeley group and
the scaling theory of Wilhelm, \etal\  Since neither the leads nor the
2DEG can be ignored, $P_{\text{tot}}(E)$ must be a convolution of
$P_{\ell}(E)$ and $P_{2D}(E)$. In that case, our own calculations
indicate that the power law behavior described in Wilhelm, \etal, does
not survive the convolution.  The resulting dependence of \gset\ on
\gtd\ does however resemble power law behavior over a limited range of
\gtd, with an exponent which can vary with temperature.  Whether charge
averaging effects are important in the Berkeley system is unclear to us
at this time.

The most significant puzzle associated with our own work is the sudden
disagreement between theory and experiment between $T=150$ and
\amount{100}{mK}.  While it is clear that the perturbative expression
for the tunneling rates in Eq.~\ref{gameq} has a low temperature bound
for applicability, it is unclear why theory and experiment disagree at
\amount{100}{mK} for \textit{all} $\gtd<6.5G_{0}$.  This disagreement,
as well as the continued reduction in current for temperatures below
\amount{100}{mK} is suggestive either of a limitation in the
environmental theory, or that some other physics is beginning to
dominate at the lowest temperatures.  Note that while \ztd\ becomes
relatively large, the impedance seen by the junctions never becomes much
larger than about \amount{1000}{\Omega} at any frequency since both
$\kappa_{1}$ and $\kappa_{2}$ are relatively small, so that we are still
in the low impedance limit $\rezt<\rqu$. Difficulties with the
environmental theory in this impedance range could be of great
importance to potential quantum computation applications. Finally, we do
not believe that Coulomb blockade physics in the 2DEG is likely to be of
importance, since the QPC conductances are still quite high even for the
smallest values of \gtd; Coulomb blockade oscillations do not begin to
appear in the 2DEG until both  QPCs have a conductance below $G_{0}$.

It is straightforward to propose experiments which could address these
issues.  The simplest would be to perform more measurements like those
in Fig.~\ref{gvsvgfig} for samples with smaller $E_{J}$.  (Such
measurements were unfortunately not performed for S1).  One could also
simplify the analysis significantly by constructing samples in which the
2DEG is selectively removed beneath the leads.  In that case, the leads
would no longer act as transmission lines, and would present a small
impedance to the tunneling electrons, so that in all probability only
\gtd\ would be of importance. We can also imagine reducing the coupling
capacitance \ctd\ so as to reduce the effective environmental impedance
and keep the S-SET current relatively large even for large \ztd.  This
could be particularly useful if the 2DEG is further confined so as to
form a quantum dot, in which case the S-SET could be used to probe the
dot impedance and energy level structure.

\begin{acknowledgments}
This research was supported at Rice by the NSF under Grant No.\
DMR-9974365 and by the Robert A. Welch foundation.  One of us (A. J. R.)
acknowledges support from the Alfred P. Sloan Foundation.  We thank K.
D. Maranowski and A. C. Gossard for providing the 2DEG material.

\end{acknowledgments}


\begin{thebibliography}{41}
\expandafter\ifx\csname natexlab\endcsname\relax\def\natexlab#1{#1}\fi
\expandafter\ifx\csname bibnamefont\endcsname\relax
  \def\bibnamefont#1{#1}\fi
\expandafter\ifx\csname bibfnamefont\endcsname\relax
  \def\bibfnamefont#1{#1}\fi
\expandafter\ifx\csname citenamefont\endcsname\relax
  \def\citenamefont#1{#1}\fi
\expandafter\ifx\csname url\endcsname\relax
  \def\url#1{\texttt{#1}}\fi
\expandafter\ifx\csname urlprefix\endcsname\relax\def\urlprefix{URL }\fi
\providecommand{\bibinfo}[2]{#2}
\providecommand{\eprint}[2][]{\url{#2}}

\bibitem[{\citenamefont{Makhlin et~al.}(1999)\citenamefont{Makhlin,
  Sch{\"{o}}n, and Shnirman}}]{Makhlin:1999}
\bibinfo{author}{\bibfnamefont{{\mbox{Yu}}.}~\bibnamefont{Makhlin}},
  \bibinfo{author}{\bibfnamefont{G.}~\bibnamefont{Sch{\"{o}}n}},
  \bibnamefont{and} \bibinfo{author}{\bibfnamefont{A.}~\bibnamefont{Shnirman}},
  \bibinfo{journal}{Nature} \textbf{\bibinfo{volume}{398}},
  \bibinfo{pages}{305} (\bibinfo{year}{1999}).

\bibitem[{\citenamefont{Nakamura et~al.}(1999)\citenamefont{Nakamura, Pashkin,
  and Tsai}}]{Nakamura:1999}
\bibinfo{author}{\bibfnamefont{Y.}~\bibnamefont{Nakamura}},
  \bibinfo{author}{\bibfnamefont{{\mbox{Yu}}.~A.} \bibnamefont{Pashkin}},
  \bibnamefont{and} \bibinfo{author}{\bibfnamefont{J.~S.} \bibnamefont{Tsai}},
  \bibinfo{journal}{Nature} \textbf{\bibinfo{volume}{398}},
  \bibinfo{pages}{786} (\bibinfo{year}{1999}).

\bibitem[{\citenamefont{Mooij et~al.}(1999)\citenamefont{Mooij, Orlando,
  Levitov, Tian, van~der Wal, and Lloyd}}]{Mooij:1999}
\bibinfo{author}{\bibfnamefont{J.~E.} \bibnamefont{Mooij}},
  \bibinfo{author}{\bibfnamefont{T.~P.} \bibnamefont{Orlando}},
  \bibinfo{author}{\bibfnamefont{L.}~\bibnamefont{Levitov}},
  \bibinfo{author}{\bibfnamefont{L.}~\bibnamefont{Tian}},
  \bibinfo{author}{\bibfnamefont{C.~H.} \bibnamefont{van~der Wal}},
  \bibnamefont{and} \bibinfo{author}{\bibfnamefont{S.}~\bibnamefont{Lloyd}},
  \bibinfo{journal}{Science} \textbf{\bibinfo{volume}{285}},
  \bibinfo{pages}{1036} (\bibinfo{year}{1999}).

\bibitem[{\citenamefont{Vion et~al.}(2002)\citenamefont{Vion, Aassime, Cottet,
  Joyez, Pothier, Urbina, Esteve, and Devoret}}]{Vion:2002}
\bibinfo{author}{\bibfnamefont{D.}~\bibnamefont{Vion}},
  \bibinfo{author}{\bibfnamefont{A.}~\bibnamefont{Aassime}},
  \bibinfo{author}{\bibfnamefont{A.}~\bibnamefont{Cottet}},
  \bibinfo{author}{\bibfnamefont{P.}~\bibnamefont{Joyez}},
  \bibinfo{author}{\bibfnamefont{H.}~\bibnamefont{Pothier}},
  \bibinfo{author}{\bibfnamefont{C.}~\bibnamefont{Urbina}},
  \bibinfo{author}{\bibfnamefont{D.}~\bibnamefont{Esteve}}, \bibnamefont{and}
  \bibinfo{author}{\bibfnamefont{M.~H.} \bibnamefont{Devoret}},
  \bibinfo{journal}{Science} \textbf{\bibinfo{volume}{296}},
  \bibinfo{pages}{886} (\bibinfo{year}{2002}).

\bibitem[{\citenamefont{Shnirman et~al.}(1997)\citenamefont{Shnirman,
  Sch{\"{o}}n, and Hermon}}]{Shnirman:1997}
\bibinfo{author}{\bibfnamefont{A.}~\bibnamefont{Shnirman}},
  \bibinfo{author}{\bibfnamefont{G.}~\bibnamefont{Sch{\"{o}}n}},
  \bibnamefont{and} \bibinfo{author}{\bibfnamefont{Z.}~\bibnamefont{Hermon}},
  \bibinfo{journal}{Phys. Rev. Lett.} \textbf{\bibinfo{volume}{79}},
  \bibinfo{pages}{2371} (\bibinfo{year}{1997}).

\bibitem[{\citenamefont{Makhlin et~al.}(2001)\citenamefont{Makhlin,
  Sch{\"{o}}n, and Shnirman}}]{Makhlin:2001a}
\bibinfo{author}{\bibfnamefont{{\mbox{Yu}}.}~\bibnamefont{Makhlin}},
  \bibinfo{author}{\bibfnamefont{G.}~\bibnamefont{Sch{\"{o}}n}},
  \bibnamefont{and} \bibinfo{author}{\bibfnamefont{A.}~\bibnamefont{Shnirman}},
  \bibinfo{journal}{Rev. Mod. Phys.} \textbf{\bibinfo{volume}{73}},
  \bibinfo{pages}{357} (\bibinfo{year}{2001}).

\bibitem[{\citenamefont{Fujisawa et~al.}(1998)\citenamefont{Fujisawa,
  Oosterkamp, van~der Wiel, Broer, Aguado, Tarucha, and
  Kouwenhoven}}]{Fujisawa:1998}
\bibinfo{author}{\bibfnamefont{T.}~\bibnamefont{Fujisawa}},
  \bibinfo{author}{\bibfnamefont{T.~H.} \bibnamefont{Oosterkamp}},
  \bibinfo{author}{\bibfnamefont{W.~G.} \bibnamefont{van~der Wiel}},
  \bibinfo{author}{\bibfnamefont{B.~W.} \bibnamefont{Broer}},
  \bibinfo{author}{\bibfnamefont{R.}~\bibnamefont{Aguado}},
  \bibinfo{author}{\bibfnamefont{S.}~\bibnamefont{Tarucha}}, \bibnamefont{and}
  \bibinfo{author}{\bibfnamefont{L.~P.} \bibnamefont{Kouwenhoven}},
  \bibinfo{journal}{Science} \textbf{\bibinfo{volume}{282}},
  \bibinfo{pages}{932} (\bibinfo{year}{1998}).

\bibitem[{\citenamefont{Aguado and Kouwenhoven}(2000)}]{Aguado:2000}
\bibinfo{author}{\bibfnamefont{R.}~\bibnamefont{Aguado}} \bibnamefont{and}
  \bibinfo{author}{\bibfnamefont{L.~P.} \bibnamefont{Kouwenhoven}},
  \bibinfo{journal}{Phys. Rev. Lett.} \textbf{\bibinfo{volume}{84}},
  \bibinfo{pages}{1986} (\bibinfo{year}{2000}).

\bibitem[{\citenamefont{Sohdhi et~al.}(1997)\citenamefont{Sohdhi, Girvin,
  Carini, and Shahar}}]{Sondhi:1997}
\bibinfo{author}{\bibfnamefont{S.~L.} \bibnamefont{Sohdhi}},
  \bibinfo{author}{\bibfnamefont{S.~M.} \bibnamefont{Girvin}},
  \bibinfo{author}{\bibfnamefont{J.~P.} \bibnamefont{Carini}},
  \bibnamefont{and} \bibinfo{author}{\bibfnamefont{D.}~\bibnamefont{Shahar}},
  \bibinfo{journal}{Rev. Mod. Phys.} \textbf{\bibinfo{volume}{69}},
  \bibinfo{pages}{315} (\bibinfo{year}{1997}).

\bibitem[{\citenamefont{Mason and Kapitulnik}(1999)}]{Mason:1999}
\bibinfo{author}{\bibfnamefont{N.}~\bibnamefont{Mason}} \bibnamefont{and}
  \bibinfo{author}{\bibfnamefont{A.}~\bibnamefont{Kapitulnik}},
  \bibinfo{journal}{Phys. Rev. Lett.} \textbf{\bibinfo{volume}{82}},
  \bibinfo{pages}{5341} (\bibinfo{year}{1999}).

\bibitem[{\citenamefont{Kapitulnik et~al.}(2001)\citenamefont{Kapitulnik,
  Mason, Kivelson, and Chakravarty}}]{Kapitulnik:2001}
\bibinfo{author}{\bibfnamefont{A.}~\bibnamefont{Kapitulnik}},
  \bibinfo{author}{\bibfnamefont{N.}~\bibnamefont{Mason}},
  \bibinfo{author}{\bibfnamefont{S.~A.} \bibnamefont{Kivelson}},
  \bibnamefont{and}
  \bibinfo{author}{\bibfnamefont{S.}~\bibnamefont{Chakravarty}},
  \bibinfo{journal}{Phys. Rev. B} \textbf{\bibinfo{volume}{63}},
  \bibinfo{pages}{125322} (\bibinfo{year}{2001}).

\bibitem[{\citenamefont{Rimberg et~al.}(1997)\citenamefont{Rimberg, Ho, Kurdak,
  Clarke, Campman, and Gossard}}]{Rimberg:1997}
\bibinfo{author}{\bibfnamefont{A.~J.} \bibnamefont{Rimberg}},
  \bibinfo{author}{\bibfnamefont{T.~R.} \bibnamefont{Ho}},
  \bibinfo{author}{\bibfnamefont{{\c{C}}.}~\bibnamefont{Kurdak}},
  \bibinfo{author}{\bibfnamefont{J.}~\bibnamefont{Clarke}},
  \bibinfo{author}{\bibfnamefont{K.~L.} \bibnamefont{Campman}},
  \bibnamefont{and} \bibinfo{author}{\bibfnamefont{A.~C.}
  \bibnamefont{Gossard}}, \bibinfo{journal}{Phys. Rev. Lett.}
  \textbf{\bibinfo{volume}{78}}, \bibinfo{pages}{2632} (\bibinfo{year}{1997}).

\bibitem[{\citenamefont{Bouchiat et~al.}(1998)\citenamefont{Bouchiat, Vion,
  Joyez, Esteve, and Devoret}}]{Bouchiat:1998}
\bibinfo{author}{\bibfnamefont{V.}~\bibnamefont{Bouchiat}},
  \bibinfo{author}{\bibfnamefont{D.}~\bibnamefont{Vion}},
  \bibinfo{author}{\bibfnamefont{P.}~\bibnamefont{Joyez}},
  \bibinfo{author}{\bibfnamefont{D.}~\bibnamefont{Esteve}}, \bibnamefont{and}
  \bibinfo{author}{\bibfnamefont{M.~H.} \bibnamefont{Devoret}},
  \bibinfo{journal}{Phys. Scripta} \textbf{\bibinfo{volume}{T76}},
  \bibinfo{pages}{165} (\bibinfo{year}{1998}).

\bibitem[{\citenamefont{Grabert and Devoret}(1992)}]{Grabert:1992}
\bibinfo{editor}{\bibfnamefont{H.}~\bibnamefont{Grabert}} \bibnamefont{and}
  \bibinfo{editor}{\bibfnamefont{M.~H.} \bibnamefont{Devoret}}, eds.,
  \emph{\bibinfo{title}{Single Charge Tunneling}} (\bibinfo{publisher}{Plenum},
  \bibinfo{address}{New York}, \bibinfo{year}{1992}).

\bibitem[{\citenamefont{Turlot et~al.}(1989)\citenamefont{Turlot, Esteve,
  Urbina, Martinis, Devoret, Linkwitz, and Grabert}}]{Turlot:1989}
\bibinfo{author}{\bibfnamefont{E.}~\bibnamefont{Turlot}},
  \bibinfo{author}{\bibfnamefont{D.}~\bibnamefont{Esteve}},
  \bibinfo{author}{\bibfnamefont{C.}~\bibnamefont{Urbina}},
  \bibinfo{author}{\bibfnamefont{J.~M.} \bibnamefont{Martinis}},
  \bibinfo{author}{\bibfnamefont{M.~H.} \bibnamefont{Devoret}},
  \bibinfo{author}{\bibfnamefont{S.}~\bibnamefont{Linkwitz}}, \bibnamefont{and}
  \bibinfo{author}{\bibfnamefont{H.}~\bibnamefont{Grabert}},
  \bibinfo{journal}{Phys. Rev. Lett.} \textbf{\bibinfo{volume}{62}},
  \bibinfo{pages}{1788} (\bibinfo{year}{1989}).

\bibitem[{\citenamefont{Kycia et~al.}(2001)\citenamefont{Kycia, Chen, Therrien,
  Kurdak, Campman, Gossard, and Clarke}}]{Kycia:2001}
\bibinfo{author}{\bibfnamefont{J.~B.} \bibnamefont{Kycia}},
  \bibinfo{author}{\bibfnamefont{J.}~\bibnamefont{Chen}},
  \bibinfo{author}{\bibfnamefont{R.}~\bibnamefont{Therrien}},
  \bibinfo{author}{\bibfnamefont{{\c{C}}.}~\bibnamefont{Kurdak}},
  \bibinfo{author}{\bibfnamefont{K.~L.} \bibnamefont{Campman}},
  \bibinfo{author}{\bibfnamefont{A.~C.} \bibnamefont{Gossard}},
  \bibnamefont{and} \bibinfo{author}{\bibfnamefont{J.}~\bibnamefont{Clarke}},
  \bibinfo{journal}{Phys. Rev. Lett.} \textbf{\bibinfo{volume}{87}},
  \bibinfo{pages}{017002} (\bibinfo{year}{2001}).

\bibitem[{\citenamefont{Lu et~al.}(2002)\citenamefont{Lu, Maranowski, and
  Rimberg}}]{Lu:2002}
\bibinfo{author}{\bibfnamefont{W.}~\bibnamefont{Lu}},
  \bibinfo{author}{\bibfnamefont{K.~D.} \bibnamefont{Maranowski}},
  \bibnamefont{and} \bibinfo{author}{\bibfnamefont{A.~J.}
  \bibnamefont{Rimberg}}, \bibinfo{journal}{Phys. Rev. B}
  \textbf{\bibinfo{volume}{65}}, \bibinfo{pages}{060501}
  (\bibinfo{year}{2002}).

\bibitem[{\citenamefont{Lu et~al.}()\citenamefont{Lu, Maranowski, and
  Rimberg}}]{Lu:2002a}
\bibinfo{author}{\bibfnamefont{W.}~\bibnamefont{Lu}},
  \bibinfo{author}{\bibfnamefont{K.~D.} \bibnamefont{Maranowski}},
  \bibnamefont{and} \bibinfo{author}{\bibfnamefont{A.~J.}
  \bibnamefont{Rimberg}}, \eprint{cond-mat/0204287}.

\bibitem[{\citenamefont{Wilhelm et~al.}(2001)\citenamefont{Wilhelm,
  Sch{\"{o}}n, and Zim{\'{a}}nyi}}]{Wilhelm:2001}
\bibinfo{author}{\bibfnamefont{F.~K.} \bibnamefont{Wilhelm}},
  \bibinfo{author}{\bibfnamefont{G.}~\bibnamefont{Sch{\"{o}}n}},
  \bibnamefont{and} \bibinfo{author}{\bibfnamefont{G.~T.}
  \bibnamefont{Zim{\'{a}}nyi}}, \bibinfo{journal}{Phys. Rev. Lett.}
  \textbf{\bibinfo{volume}{87}}, \bibinfo{pages}{136802}
  (\bibinfo{year}{2001}).

\bibitem[{\citenamefont{Lu et~al.}(2000)\citenamefont{Lu, Rimberg, Maranowski,
  and Gossard}}]{Lu:2000}
\bibinfo{author}{\bibfnamefont{W.}~\bibnamefont{Lu}},
  \bibinfo{author}{\bibfnamefont{A.~J.} \bibnamefont{Rimberg}},
  \bibinfo{author}{\bibfnamefont{K.~D.} \bibnamefont{Maranowski}},
  \bibnamefont{and} \bibinfo{author}{\bibfnamefont{A.~C.}
  \bibnamefont{Gossard}}, \bibinfo{journal}{Appl. Phys. Lett.}
  \textbf{\bibinfo{volume}{77}}, \bibinfo{pages}{2746} (\bibinfo{year}{2000}).

\bibitem[{\citenamefont{Ingold et~al.}(1991)\citenamefont{Ingold, Wyrowski, and
  Grabert}}]{Ingold:1991}
\bibinfo{author}{\bibfnamefont{G.-L.} \bibnamefont{Ingold}},
  \bibinfo{author}{\bibfnamefont{P.}~\bibnamefont{Wyrowski}}, \bibnamefont{and}
  \bibinfo{author}{\bibfnamefont{H.}~\bibnamefont{Grabert}},
  \bibinfo{journal}{Z. Phys. B} \textbf{\bibinfo{volume}{85}},
  \bibinfo{pages}{443} (\bibinfo{year}{1991}).

\bibitem[{\citenamefont{Odintsov et~al.}(1991)\citenamefont{Odintsov, Falci,
  and Sch{\"{o}}n}}]{Odintsov:1991}
\bibinfo{author}{\bibfnamefont{A.~A.} \bibnamefont{Odintsov}},
  \bibinfo{author}{\bibfnamefont{G.}~\bibnamefont{Falci}}, \bibnamefont{and}
  \bibinfo{author}{\bibfnamefont{G.}~\bibnamefont{Sch{\"{o}}n}},
  \bibinfo{journal}{Phys. Rev. B} \textbf{\bibinfo{volume}{44}},
  \bibinfo{pages}{13~089} (\bibinfo{year}{1991}).

\bibitem[{\citenamefont{Ambegaokar and Baratoff}(1963)}]{Ambegaokar:1963}
\bibinfo{author}{\bibfnamefont{V.}~\bibnamefont{Ambegaokar}} \bibnamefont{and}
  \bibinfo{author}{\bibfnamefont{A.}~\bibnamefont{Baratoff}},
  \bibinfo{journal}{Phys. Rev. Lett.} \textbf{\bibinfo{volume}{10}},
  \bibinfo{pages}{486} (\bibinfo{year}{1963}).

\bibitem[{\citenamefont{Ingold and Nazarov}(1992)}]{Ingold:1992}
\bibinfo{author}{\bibfnamefont{G.-L.} \bibnamefont{Ingold}} \bibnamefont{and}
  \bibinfo{author}{\bibfnamefont{{\mbox{Yu}}.~V.} \bibnamefont{Nazarov}}, in
  \cite{Grabert:1992}, pp. \bibinfo{pages}{21--107}.

\bibitem[{\citenamefont{Sch{\"{o}}n}(1998)}]{Schon:1998a}
\bibinfo{author}{\bibfnamefont{G.}~\bibnamefont{Sch{\"{o}}n}}, in
  \emph{\bibinfo{booktitle}{Quantum Transport and Dissipation}}
  (\bibinfo{publisher}{Wiley-VCH}, \bibinfo{address}{Weinheim, Germany},
  \bibinfo{year}{1998}), pp. \bibinfo{pages}{149--212}.

\bibitem[{\citenamefont{Averin et~al.}(1990)\citenamefont{Averin, Nazarov, and
  Odintsov}}]{Averin:1990}
\bibinfo{author}{\bibfnamefont{D.~V.} \bibnamefont{Averin}},
  \bibinfo{author}{\bibfnamefont{{\mbox{Yu}}.~V.} \bibnamefont{Nazarov}},
  \bibnamefont{and} \bibinfo{author}{\bibfnamefont{A.~A.}
  \bibnamefont{Odintsov}}, \bibinfo{journal}{Physica B}
  \textbf{\bibinfo{volume}{165{\&}166}}, \bibinfo{pages}{945}
  (\bibinfo{year}{1990}).

\bibitem[{\citenamefont{Grabert et~al.}(1991)\citenamefont{Grabert, Ingold,
  Devoret, Est{\`{e}}ve, Pothier, and Urbina}}]{Grabert:1991}
\bibinfo{author}{\bibfnamefont{H.}~\bibnamefont{Grabert}},
  \bibinfo{author}{\bibfnamefont{G.-L.} \bibnamefont{Ingold}},
  \bibinfo{author}{\bibfnamefont{M.~H.} \bibnamefont{Devoret}},
  \bibinfo{author}{\bibfnamefont{D.}~\bibnamefont{Est{\`{e}}ve}},
  \bibinfo{author}{\bibfnamefont{H.}~\bibnamefont{Pothier}}, \bibnamefont{and}
  \bibinfo{author}{\bibfnamefont{C.}~\bibnamefont{Urbina}},
  \bibinfo{journal}{Z. Phys. B} \textbf{\bibinfo{volume}{84}},
  \bibinfo{pages}{143} (\bibinfo{year}{1991}).

\bibitem[{\citenamefont{Pozar}(1998)}]{Pozar:1998}
\bibinfo{author}{\bibfnamefont{D.~M.} \bibnamefont{Pozar}},
  \emph{\bibinfo{title}{Microwave Engineering, 2nd Ed.}}
  (\bibinfo{publisher}{John Wiley \& Sons}, \bibinfo{address}{New York},
  \bibinfo{year}{1998}).

\bibitem[{\citenamefont{Kuzmin et~al.}(1991)\citenamefont{Kuzmin, Nazarov,
  Haviland, Delsing, and Claeson}}]{Kuzmin:1991}
\bibinfo{author}{\bibfnamefont{L.~S.} \bibnamefont{Kuzmin}},
  \bibinfo{author}{\bibfnamefont{{\mbox{Yu}}.~V.} \bibnamefont{Nazarov}},
  \bibinfo{author}{\bibfnamefont{D.~B.} \bibnamefont{Haviland}},
  \bibinfo{author}{\bibfnamefont{P.}~\bibnamefont{Delsing}}, \bibnamefont{and}
  \bibinfo{author}{\bibfnamefont{T.}~\bibnamefont{Claeson}},
  \bibinfo{journal}{Phys. Rev. Lett.} \textbf{\bibinfo{volume}{67}},
  \bibinfo{pages}{1161} (\bibinfo{year}{1991}).

\bibitem[{\citenamefont{Grabert et~al.}(1998)\citenamefont{Grabert, Ingold, and
  Paul}}]{Grabert:1998}
\bibinfo{author}{\bibfnamefont{H.}~\bibnamefont{Grabert}},
  \bibinfo{author}{\bibfnamefont{G.-L.} \bibnamefont{Ingold}},
  \bibnamefont{and} \bibinfo{author}{\bibfnamefont{B.}~\bibnamefont{Paul}},
  \bibinfo{journal}{Europhys. Lett.} \textbf{\bibinfo{volume}{44}},
  \bibinfo{pages}{360} (\bibinfo{year}{1998}).

\bibitem[{\citenamefont{Grabert and Ingold}(1999)}]{Grabert:1999}
\bibinfo{author}{\bibfnamefont{H.}~\bibnamefont{Grabert}} \bibnamefont{and}
  \bibinfo{author}{\bibfnamefont{G.-L.} \bibnamefont{Ingold}},
  \bibinfo{journal}{Supperlattices and Microstruct.}
  \textbf{\bibinfo{volume}{25}}, \bibinfo{pages}{915} (\bibinfo{year}{1999}).

\bibitem[{\citenamefont{Collin}(1992)}]{Collin:1992}
\bibinfo{author}{\bibfnamefont{R.~E.} \bibnamefont{Collin}},
  \emph{\bibinfo{title}{Foundations for Microwave Engineering}}
  (\bibinfo{publisher}{McGraw-Hill}, \bibinfo{address}{New York},
  \bibinfo{year}{1992}).

\bibitem[{\citenamefont{Visscher et~al.}(1995)\citenamefont{Visscher, Verbrugh,
  Lindeman, Hadley, and Mooij}}]{Visscher:1995}
\bibinfo{author}{\bibfnamefont{E.~H.} \bibnamefont{Visscher}},
  \bibinfo{author}{\bibfnamefont{S.~M.} \bibnamefont{Verbrugh}},
  \bibinfo{author}{\bibfnamefont{J.}~\bibnamefont{Lindeman}},
  \bibinfo{author}{\bibfnamefont{P.}~\bibnamefont{Hadley}}, \bibnamefont{and}
  \bibinfo{author}{\bibfnamefont{J.~E.} \bibnamefont{Mooij}},
  \bibinfo{journal}{Appl. Phys. Lett.} \textbf{\bibinfo{volume}{66}},
  \bibinfo{pages}{305} (\bibinfo{year}{1995}).

\bibitem[{\citenamefont{Krupenin et~al.}(1998)\citenamefont{Krupenin, Presnov,
  Savvateev, Scherer, Zorin, and Niemeyer}}]{Krupenin:1998}
\bibinfo{author}{\bibfnamefont{V.~A.} \bibnamefont{Krupenin}},
  \bibinfo{author}{\bibfnamefont{D.~E.} \bibnamefont{Presnov}},
  \bibinfo{author}{\bibfnamefont{M.~N.} \bibnamefont{Savvateev}},
  \bibinfo{author}{\bibfnamefont{H.}~\bibnamefont{Scherer}},
  \bibinfo{author}{\bibfnamefont{A.~B.} \bibnamefont{Zorin}}, \bibnamefont{and}
  \bibinfo{author}{\bibfnamefont{J.}~\bibnamefont{Niemeyer}},
  \bibinfo{journal}{J. Appl. Phys.} \textbf{\bibinfo{volume}{84}},
  \bibinfo{pages}{3212} (\bibinfo{year}{1998}).

\bibitem[{\citenamefont{Kogan}(1996)}]{Kogan:1996}
\bibinfo{author}{\bibfnamefont{{\mbox{Sh}}.}~\bibnamefont{Kogan}},
  \emph{\bibinfo{title}{Electronic Noise and Fluctuations in Solids}}
  (\bibinfo{publisher}{Cambridge University Press},
  \bibinfo{address}{Cambridge, United Kingdom}, \bibinfo{year}{1996}).

\bibitem[{\citenamefont{Eiles and Martinis}(1994)}]{Eiles:1994}
\bibinfo{author}{\bibfnamefont{T.~M.} \bibnamefont{Eiles}} \bibnamefont{and}
  \bibinfo{author}{\bibfnamefont{J.~M.} \bibnamefont{Martinis}},
  \bibinfo{journal}{Phys. Rev. B} \textbf{\bibinfo{volume}{50}},
  \bibinfo{pages}{627} (\bibinfo{year}{1994}).

\bibitem[{\citenamefont{Tuominen et~al.}(1992)\citenamefont{Tuominen,
  Hergenrother, Tighe, and Tinkham}}]{Tuominen:1992}
\bibinfo{author}{\bibfnamefont{M.~T.} \bibnamefont{Tuominen}},
  \bibinfo{author}{\bibfnamefont{J.~M.} \bibnamefont{Hergenrother}},
  \bibinfo{author}{\bibfnamefont{T.~S.} \bibnamefont{Tighe}}, \bibnamefont{and}
  \bibinfo{author}{\bibfnamefont{M.}~\bibnamefont{Tinkham}},
  \bibinfo{journal}{Phys. Rev. Lett.} \textbf{\bibinfo{volume}{69}},
  \bibinfo{pages}{1997} (\bibinfo{year}{1992}).

\bibitem[{\citenamefont{Eiles et~al.}(1993)\citenamefont{Eiles, Martinis, and
  Devoret}}]{Eiles:1993}
\bibinfo{author}{\bibfnamefont{T.~M.} \bibnamefont{Eiles}},
  \bibinfo{author}{\bibfnamefont{J.~M.} \bibnamefont{Martinis}},
  \bibnamefont{and} \bibinfo{author}{\bibfnamefont{M.~H.}
  \bibnamefont{Devoret}}, \bibinfo{journal}{Phys. Rev. Lett.}
  \textbf{\bibinfo{volume}{70}}, \bibinfo{pages}{1862} (\bibinfo{year}{1993}).

\bibitem[{\citenamefont{Amar et~al.}(1994)\citenamefont{Amar, Song, Lobb, and
  Wellstood}}]{Amar:1994}
\bibinfo{author}{\bibfnamefont{A.}~\bibnamefont{Amar}},
  \bibinfo{author}{\bibfnamefont{D.}~\bibnamefont{Song}},
  \bibinfo{author}{\bibfnamefont{C.~J.} \bibnamefont{Lobb}}, \bibnamefont{and}
  \bibinfo{author}{\bibfnamefont{F.~C.} \bibnamefont{Wellstood}},
  \bibinfo{journal}{Phys. Rev. Lett.} \textbf{\bibinfo{volume}{72}},
  \bibinfo{pages}{3234} (\bibinfo{year}{1994}).

\bibitem[{\citenamefont{Joyez et~al.}(1994)\citenamefont{Joyez, Lafarge,
  Filipe, Esteve, and Devoret}}]{Joyez:1994}
\bibinfo{author}{\bibfnamefont{P.}~\bibnamefont{Joyez}},
  \bibinfo{author}{\bibfnamefont{P.}~\bibnamefont{Lafarge}},
  \bibinfo{author}{\bibfnamefont{A.}~\bibnamefont{Filipe}},
  \bibinfo{author}{\bibfnamefont{D.}~\bibnamefont{Esteve}}, \bibnamefont{and}
  \bibinfo{author}{\bibfnamefont{M.~H.} \bibnamefont{Devoret}},
  \bibinfo{journal}{Phys. Rev. Lett.} \textbf{\bibinfo{volume}{72}},
  \bibinfo{pages}{2458} (\bibinfo{year}{1994}).

\bibitem[{\citenamefont{Covington et~al.}(2000)\citenamefont{Covington, Keller,
  Kautz, and Martinis}}]{Covington:2000}
\bibinfo{author}{\bibfnamefont{M.}~\bibnamefont{Covington}},
  \bibinfo{author}{\bibfnamefont{M.~W.} \bibnamefont{Keller}},
  \bibinfo{author}{\bibfnamefont{R.~L.} \bibnamefont{Kautz}}, \bibnamefont{and}
  \bibinfo{author}{\bibfnamefont{J.~M.} \bibnamefont{Martinis}},
  \bibinfo{journal}{Phys. Rev. Lett.} \textbf{\bibinfo{volume}{84}},
  \bibinfo{pages}{5192} (\bibinfo{year}{2000}).

\end{thebibliography}

\end{document}